\documentclass{jaa}
\usepackage{float,lscape}
\usepackage{geometry}
\usepackage{xspace}	
\usepackage{graphicx}
\usepackage{lscape}
\usepackage{threeparttable}
\usepackage{amsmath}
\usepackage[utf8]{inputenc}
\usepackage{longtable}
\usepackage{natbib}
\usepackage{multirow}
\usepackage[Table,xcdraw]{xcolor}
\usepackage{rotating}
\usepackage[nodisplayskipstretch]{setspace}
\usepackage{tabularx}
\usepackage[labelfont=bf]{caption} 
\usepackage{ragged2e}
\usepackage{lipsum}

\usepackage{hyperref}
\hypersetup{backref=true,
    pagebackref=true,
    hyperindex=true,
    colorlinks=true,
    breaklinks=true,
    urlcolor= black,
    linkcolor= blue,
    bookmarks=false,
    bookmarksopen=false,
    filecolor=black,
    citecolor=blue,
    linkbordercolor=blue
}

\begin{document}

\title{Photometric calibrations and characterization of the 4K$\times$4K CCD Imager, the first-light axial port instrument for the 3.6m DOT}

\author{Amit Kumar\textsuperscript{1,2,*}, S. B. Pandey\textsuperscript{1}, Avinash Singh\textsuperscript{1,3}, R. K. S. Yadav\textsuperscript{1}, B. K. Reddy\textsuperscript{1}, N. Nanjappa\textsuperscript{1}, S. Yadav\textsuperscript{1}, and R. Srinivasan\textsuperscript{4}}
\affilOne{\textsuperscript{1}Aryabhatta Research Institute of Observational Sciences, Manora Peak, Nainital 263002, Uttarakhand, India\\}
\affilTwo{\textsuperscript{2}School of Studies in Physics and Astrophysics, Pandit Ravishankar Shukla University, Raipur 492010, Chhattisgarh, India.\\}
\affilThree{\textsuperscript{3}Hiroshima Astrophysical Science Center, Hiroshima University, Higashi-Hiroshima, Hiroshima 739-8526, Japan.\\}
\affilFour{\textsuperscript{4}Vemana Institute of Technology, Mahayogi Vemana Road, Bangalore, 560034.\\}

\twocolumn[{

\maketitle

\corres{amit@aries.res.in, amitkundu515@gmail.com, shashi@aries.res.in}

\msinfo{2021}{2021}

\begin{abstract}
In the present work, recent characterization results of the 4K$\times$4K CCD Imager (a first light instrument of the 3.6m Devasthal Optical Telescope; DOT) and photometric calibrations are discussed, along with measurements of the extinction coefficients and sky brightness values at the location of the 3.6m DOT site based on the imaging data taken between 2016 to 2021. For the 4K$\times$4K CCD Imager, all given combinations of gains (1, 2, 3, 5, and 10 e$^-$/ADU) and readout noise values for the three readout speeds (100 kHz, 500 kHz, and 1 MHz) are verified using the sky flats and bias frames taken during early 2021; measured values resemble well with the theoretical ones. Using color-color and color-magnitude transformation equations, color coefficients ($\alpha$) and zero-points ($\beta$) are determined to constrain and examine their long-term consistencies and any possible evolution based on $UBVRI$ observations of several Landolt standard fields observed during 2016--2021. Our present analysis exhibits consistency among estimated $\alpha$ values within the 1$\sigma$ and does not show any noticeable trend with time. We also found that the photometric errors and limiting magnitudes computed using the CCD Imager data follow the simulated ones published earlier. The average extinction coefficients, their seasonal variations, and zenith night-sky brightness values for the moon-less nights for all ten Bessell and SDSS filters are also estimated and found comparable to those reported for other good astronomical sites.
\end{abstract}

}]

\doinum{}
\artcitid{\#}
\volnum{000}
\year{2021}
\pgrange{1--}
\setcounter{page}{1}
\lp{16}

\section{Introduction}

The Charged Coupled Devices (CCDs) are digital photo-detectors extensively used in optical-infrared observational astronomy, primarily working on the principle of the photoelectric effect. Based on the proposed scientific goals, the manufacturers could provide a range of various parameters of CCDs (gain, readout speed, readout noise, etc.) to make the best use of the dynamic range \citep{Howell2006}. It is always recommended to verify given CCD parameters observationally to increase observations' accuracy and control possible degradation of the CCD electronics with time. The present work discusses the characterization results of the STA-4150A 4K$\times$4K CCD Imager mounted at the axial port of the 3.6m Devasthal Optical Telescope (DOT; \citealt{Pandey2016a, Kumar2018, Pandey2018, Sagar2019, Kumar2021a}) based on the data accumulated between early 2016 to early 2021. In this work, verification of CCD parameters like gain and readout noise (RN) are performed for all given combinations of readout speeds and verification of bias stability. Preliminary characterization results for the 4K$\times$4K CCD Imager based on the data collected during 2016-2017 are presented in \citealt{Pandey2018} (hereafter \textcolor{blue}{P18}\href{https://ui.adsabs.harvard.edu/abs/2018BSRSL..87...42P/abstract}).

It is also known that the instrumental magnitudes obtained using the raw data of the CCD detectors are always required to be converted to the standard photometric systems in order to compare with those obtained from other instruments. This calibration can be done using the standard color-color or color-magnitude transformation equations for a given photometric system \citep{Romanishin2002}. CCD observations under good photometric conditions are recommended to use for a given photometric system to estimate the color coefficients and zero points. In this study, we estimated the values of color coefficients and zero points for different filters (Bessell $UBVRI$) and investigated their temporal evolution over a period of nearly five years.

Apart from the aspects mentioned above, an excellent astronomical site also needs to be characterized for different atmospheric and geographical conditions (e.g., atmospheric extinction, night sky brightness, etc.) to estimate the real brightness of observed celestial objects. 

Atmospheric extinction is one of the critical constituents which affects ground-based astronomical observations by attenuation of the light from celestial objects via scattering/absorbing photons by air molecules when it passes through the Earth's atmosphere \citep{Hayes1975}. \cite{Mohan1999} calculated the extinction coefficient in Bessell $UBVRI$ filters for a site near the 3.6m DOT, based on the data obtained during 1998--1999. Night sky brightness is another crucial factor in characterizing an astronomical site. Sky brightness values for the 3.6m DOT site in some of the optical bands are reported recently by \cite{Sagar2020}, whereas for the NIR $JHK$ bands are published by \citet{Baug2018}. As a part of the present analysis, we constrained atmospheric extinction and night sky brightness values in Bessell $UBVRI$ and SDSS $ugriz$ broadband filters for the 3.6m DOT site using the 4K$\times$4K CCD imaging data obtained between 2016 to 2021 (covering different seasons).

The paper is organized as follows. Section~\ref{sec:observations} discusses the observations and data reduction. Section~\ref{sec:Imager} gives an overview of the 4K$\times$4K CCD Imager and the ten optical broadband filters. CCD parameters like gain, $RN$, and bias stability are verified in Section~\ref{sec:bias_G_RN}. Extinction coefficients in Bessell and SDSS filters are presented in Section~\ref{sec:extinction}. Section~\ref{sec:photo_calib} describes about the photometric calibrations in detail. Estimations of the night sky brightness values are discussed in Section~\ref{sec:sky_brightness}. We conclude our results in Section~\ref{sec:result}.

\section{Observations and data reduction}
\label{sec:observations}

This study uses the characterization and calibration data obtained using the STA4150A 4K$\times$4K CCD Imager\footnote{\url{https://www.aries.res.in/sites/default/files/files/3.6-DOT/imager.pdf}} mounted at the axial port of the 3.6m DOT on several occasions between March 2016 and February 2021. For characterization purposes, bias and sky-flat frames were taken in possible combinations of the readout speeds (100 kHz, 500 kHz, and 1 MHz) and gain values (1, 2, 3, 5, and 10 e$^-$/ADU) mostly in 2$\times$2 binning and single readout mode, spread over more than a dozen of nights during early 2021. To check the bias stability of the CCD Imager, we continuously observed bias frames for at least ten hours for each readout speed of 100 kHz, 500 kHz, and 1 MHz on different nights. We also observed multiple flat frames with varying mean counts within the linearity region during several nights to cross-verify the gain and $RN$ values. The data reduction and analysis for characterization purposes were made using self-developed Python scripts along with standard IRAF routines like {\tt imstat} and {\tt imarith}. The cosmic-rays were also removed from all the bias/sky-flat frames used in the present analysis using the IRAF sub-routine {\tt cosmicrays}.

For photometric calibrations in ten broadband filters, i.e., Bessell $UBVRI$/SDSS $ugriz$ and characterization of the Devasthal site (Latitude: 29$^\circ$ 23$'$ North; Longitude: 79$^\circ$ 41$'$ East; Altitude: 2540 m) in terms of extinction coefficient measurements and estimation of night-sky brightness based on observations of many Landolt standard field \citep[PG~0918, PG~2213, PG~0231, PG~1633, SA~104, PG~1323, SA~110, PG~1047, PG~1525, SA~111, SA~113, SA~98, PG~2331, SA~92, PG~1657;][]{Landolt1992} are also presented as a part of this paper. The Landolt standard fields mentioned above have stars with $V$-band magnitudes range of $\sim$10.07 to 16.40 mag and a $B-V$ color range of $\sim-$0.33 to +1.45 mag. The complete observation log of the Landolt standard fields used in the present study is tabulated in Table~\ref{tab:log_table}. The observations of Landolt standard fields tabulated in Table~\ref{tab:log_table} were taken within the airmass range from zenith to $\sim$4.4, with a range of lunar phases and Full-Width Half-Maximum (FWHM) from $\sim$0.45 to 2 arcsec (in $V$-band). During these observations, reflectively of the primary mirror ($M1$) of the 3.6m DOT varied between $\sim$49 and 85\%. We took most of the data in a 2$\times$2 binning, gain = 5 e$^-$/ADU, and in single readout mode with a readout speed of 1 MHz.

The pre-processing of raw data was done through standard procedures by applying bias correction, flat fielding, and cosmic ray removal using IRAF routines and python-based scripts hosted on \textsc{RedPipe} \citep{2021redpipe}. Wherever required, multiple frames in a single band observed on the same night were aligned using the python-based code {\tt alipy} and then median-combined utilizing the IRAF sub-routine {\tt IMARITH} to attain a better signal-to-noise ratio. All the photometry and standard calibration procedures were carried out using self-developed Python scripts hosted on \textsc{RedPipe} \citep{2021redpipe}. For extinction coefficient measurements, we observed six sets of Landolt standard fields \citep{Landolt1992} in $UBVRI$ / $ugriz$ filters, covering an airmass range of $\sim$1.3 to 4.4 using the 4K$\times$4K CCD Imager over several nights from 2017 to 2021. 32 Landolt standard fields observed in 5 Bessell filters ($UBVRI$) distributed over 20 different nights between 2016 March and 2021 February are used for the photometric calibrations. Whereas all Landolt standard fields tabulated in Table~\ref{tab:log_table} are used for night sky brightness estimation irrespective of spectral coverage.

It is also worth mentioning here that the observations used in the present analysis were performed randomly (whenever the CCD Imager was mounted and the sky was clear) in a diverse set of observing conditions, e.g., airmass, humidity, moon angle, moon distance, $M1$ reflectively, etc. As a result, a diverse range of FWHM was obtained for the dataset used in this analysis. Hence, for reference, we have plotted the FWHM values (in arcsec) as obtained for $V$-band (within airmass range $\sim$1.1--1.5) as a function of time in Figure~\ref{fig:fwhm_three_cycle}. The best FWHM value of $\sim$0.43 arcsec ($\sim$2.25 pixels in 2$\times$2 binned mode) for a point object in the field of SN~2017egm \citep{Nicholl2017} based on the observations taken on 2020 March 17 is also reported as an example (see $r$-band stellar image in Figure~\ref{fig:fwhm_best}). We have also shown the contour and radial profiles of this stellar image in Figure~\ref{fig:fwhm_best}. It is also worth mentioning that using the observations with TIFR NIR Imaging Camera-II (TIRCAM2) at 3.6m DOT on 2017 October, FWHM of $\sim$ 0.45 arcsec in $K$-band was reported by \cite{Baug2018}.

\section{4K$\times$4K CCD imaging camera: a first-light back-end instrument\label{sec:Imager}}

\begin{figure}
\centering
\includegraphics[angle=0,scale=0.28]{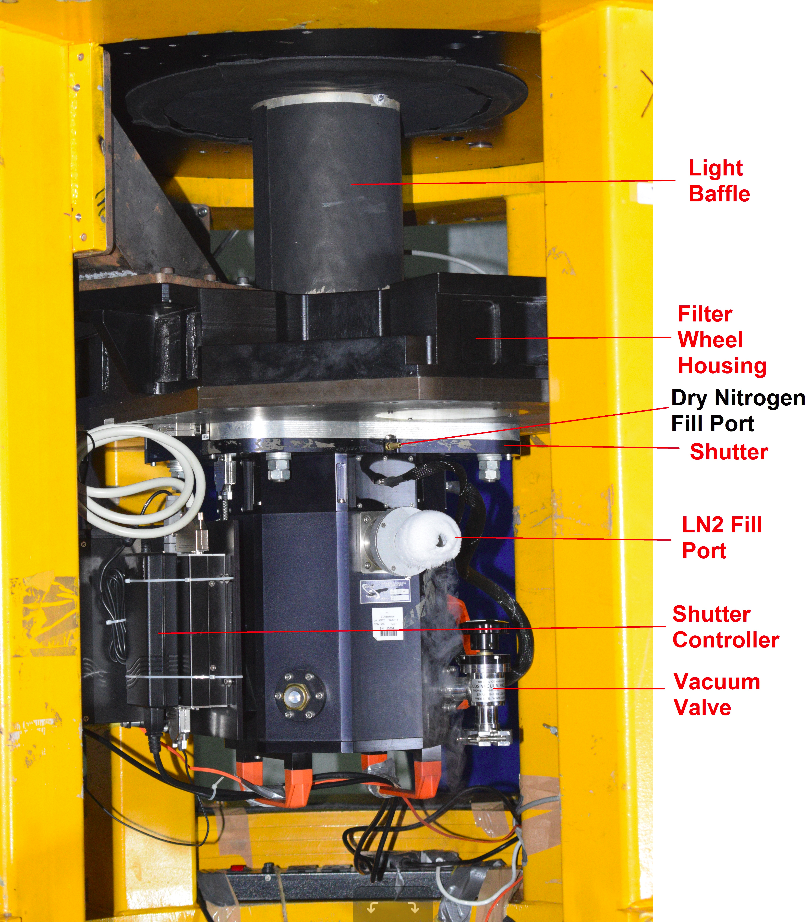}
\caption{The fully assembled 4K$\times$4K CCD Imager and the new filter wheel assembly and cylindrical light baffle as mounted at the axial port of the 3.6m DOT in early 2021. Major sub-components of the first light instrument are also designated.}
\label{fig:new_setup}
\end{figure}

The 4K$\times$4K CCD Imager is the first light instrument designed to be mounted at the axial port of the 3.6m DOT (see Figure~\ref{fig:new_setup}). The Imager (including the chip and the controller combinations) was designed and developed by Semiconductor Technology Associates, Inc. (STA\footnote{http://www.sta-inc.net}). The Imager has a blue-enhanced back illuminated CCD chip having 4096$\times$4096 pixels (15 $\mu$m pixel size). The CCD Dewar is cooled with Liquid Nitrogen (LN$_2$) at pressure $<$ 5 millitorr to attain a temperature of --120$^\circ C$ with an LN$_2$ hold time of $\sim$14-16 hours. The dark current of the chip is $\sim$0.0005 e$^-$ pixel$^{-1}$ s$^{-1}$ at --120$^\circ C$ (\textcolor{blue}{P18}\href{https://ui.adsabs.harvard.edu/abs/2018BSRSL..87...42P/abstract}). The CCD can be operated with one of the three integration modes, including Non-Multi Pinned Phase (Non-MPP), MPP and, Clocked Anti-blooming (CAB). The Non-MPP mode has a higher full-well capacity (265K e$^-$) and was chosen as the most desired mode of operation during our analysis. The CCD has a 16-bit analog-to-digital converter (ADC) and can thus represent 65535 analog-to-digital unit (ADU) counts. There are three different readout speeds (100 kHz, 500 kHz, and 1 MHz) and five gain values (1, 2, 3, 5, and 10 e$^-$/ADU) to cover a range of observations utilizing the full dynamic range of the camera. In addition, three different binning modes (2$\times$2, 3$\times$3, and 4$\times$4) are also available. Higher binning modes have a lower resolution but faster readout speeds and sensitivity. Readout times @ 100 kHz, 500 kHz, and 1 MHz are $\sim$41.9, 8.4, and 4.2 seconds, respectively, for 2$\times$2 binning in a single readout mode. This could further be reduced by a factor of 4 using the quad mode readout. The 4K$\times$4K CCD Imager at the 3.6m DOT provides a plate scale of $\sim$6.4$''$ mm$^{-1}$, enabling observers to get an image with a field of view of $\approx$6.5$'$ $\times$ 6.5$'$. The preliminary study of 4K$\times$4K CCD Imager based on the characterization data obtained in the lab and open sky in 2016--2017, filter-wheel automation etc., are presented in \textcolor{blue}{P18}\href{https://ui.adsabs.harvard.edu/abs/2018BSRSL..87...42P/abstract}. The detailed observing manuals for the Imager, along with mounting procedures and pointing models, etc., were also prepared and provided for general users of the 3.6m DOT\footnote{\url{https://www.aries.res.in/facilities/astronomical-telescopes/360cm-telescope/Instruments}}.

\begin{figure}
\centering
\includegraphics[angle=0,scale=0.60]{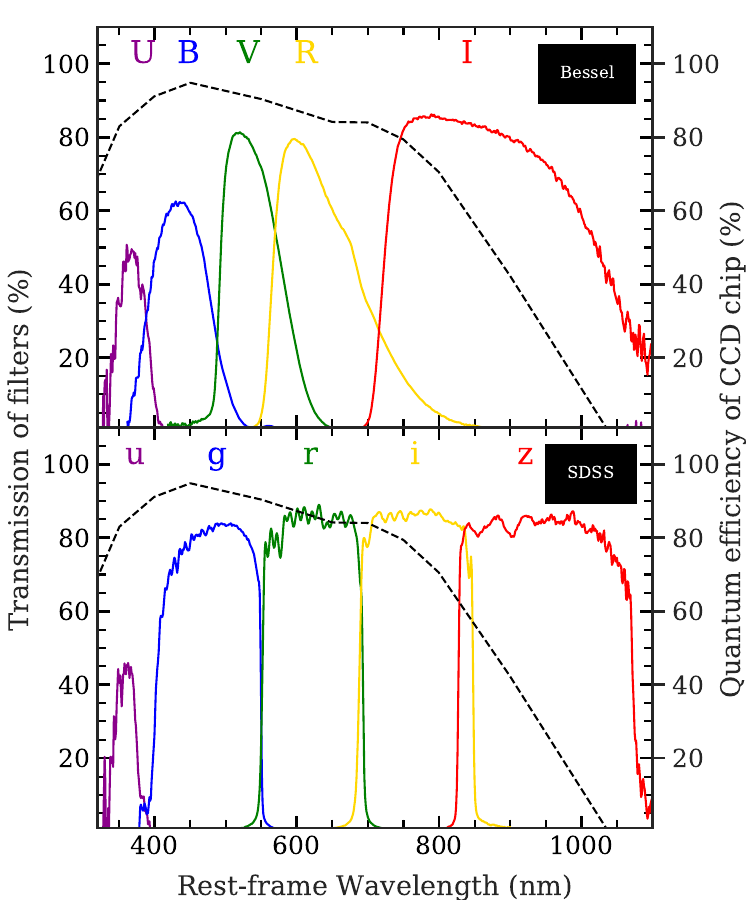}
\caption{Transmissions curves for the Bessell ($UBVRI$) and SDSS ($ugriz$) sets of filters and the quantum efficiency curve of the STA CCD chip (black dashed line) used with the 4K$\times$4K CCD Imager.}
\label{fig:filters}
\end{figure}

In 2018, a new filter wheel housing and two filter wheels were manufactured and assembled with the CCD Imager. To stop the light leakage within the Imager setup, a new cylindrical stray light baffle was designed and manufactured at ARIES and was mounted with the CCD Imager in 2020 (see Figure~\ref{fig:new_setup}). A spare CCD controller, shutter controller, power supply, and connecting wires were also procured from STA as a backup. Bessell ($UBVRI$) and SDSS ($ugriz$) broadband filters \citep{Bessell2005} with $\sim$90 mm$^2$ each in size along with 125-mm Bonn shutter are available with the CCD Imager to perform imaging of a variety of scientific objects. This set of ten broadband filters can cover a spectral range of $\approx$ 3600--10,000 \AA. Based on the tests performed in 2017, the transmission values for each of the ten broadband filters were reported in \textcolor{blue}{P18}\href{https://ui.adsabs.harvard.edu/abs/2018BSRSL..87...42P/abstract}~(U $\sim$62, B $\sim$70, V $\sim$80, R $\sim$88, I $\sim$80, u $\sim$63, g $\sim$88, r $\sim$85, i $\sim$87, and z $\sim$88\%). Transmission curves of the Bessell and SDSS filters were measured again in 2020 in the lab using a setup with a tungsten halogen lamp as a light-emitting source, a filter holder, and a table-top handheld spectrograph as a light-receiver sensitive between $\sim$4000 to 10,000 \AA. The measured transmission values are U $\sim$49, B $\sim$62, V $\sim$80, R $\sim$79, I $\sim$80, u $\sim$45, g $\sim$83, r $\sim$85, i $\sim$86, and z $\sim$85\%. The filter transmission curves along with the quantum efficiency of the STA CCD chip in per cent as a function of wavelength (black dashed line) are plotted in Figure~\ref{fig:filters}.

\section{Characterization of the 4K$\times$4K CCD Imager\label{sec:bias_G_RN}}

For the 4K$\times$4K CCD Imager, preliminary verification of gain, readout noise, linearity, etc., to a certain extent, were presented by \textcolor{blue}{P18}\href{https://ui.adsabs.harvard.edu/abs/2018BSRSL..87...42P/abstract}~based on the data taken during 2016-2017. The data used in the present study for characterization of the CCD Imager are taken in clear sky conditions while it was mounted at the axial port of the 3.6m DOT during early 2021.

\subsection{Bias stability}
\label{sec:bias_stability}

\begin{figure*}
\centering
\includegraphics[angle=0,scale=1]{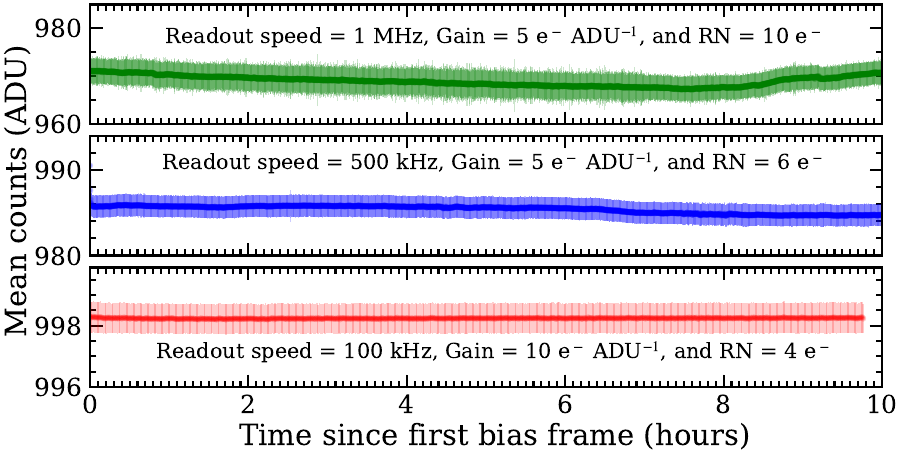}
\caption{The mean counts (ADUs) along with standard deviation (1$\sigma$) versus time of observations for the bias-frames acquired continuously nearly for 10 hours (at constant CCD temperature of --120$^\circ C$) at readout speeds 1 MHz (gain = 5 e$^-$/ADU), 500 kHz (gain = 5 e$^-$/ADU), and 100 kHz (gain = 10 e$^-$/ADU), respectively are plotted to show the stable performance of the CCD Imager. The 1$\sigma$ deviations are well within the expected readout noise values even for higher readout speeds.}
\label{fig:bias_stability}
\end{figure*}

Bias frames are the zero-second exposure readout maps associated with the CCD electronics. These frames can also be acquired from the over-scan regions of the CCD chip. Bias frames from a CCD should be stable throughout the operation to obtain accurate photometry (the mean counts should not change during observation). To check the bias stability, we observed bias frames continuously at least for ten hours at readout speeds of 1 MHz (gain = 5 e$^-$/ADU), 500 kHz (gain = 5 e$^-$/ADU), and 100 kHz (gain = 10 e$^-$/ADU) on several occasions. Mean counts and standard deviations were calculated at several locations of the chip by taking patches of at least 100$\times$100 pixels. Results thus derived for readout speeds of 1 MHz, 500 kHz, and 100 kHz are shown in the upper, middle, and lower panels of Figure~\ref{fig:bias_stability}, respectively. The y- and x-ordinates of Figure~\ref{fig:bias_stability} represent mean counts in ADU and the time in hours since the first bias frame, whereas standard deviation values represent 1$\sigma$ errors in the mean counts (shown with the shaded regions in Figure~\ref{fig:bias_stability}). Based on the observed data, the bias stability (within errors) was found as per the given specifications of the CCD, i.e., for a constant CCD temperature of --120$^\circ C$ for more than 10 hours (or LN2 hold time).

\subsection{Gain and Readout noise}

The gain of a CCD represents the number of collected electrons (or photons) to produce one ADU and is generally expressed in terms of e$^-$/ADU. The observers can choose the system gain value based on proposed scientific goals. System gain is selected to compromise between high digitization noise and loss of full well depth. Choice of lower gain results in lower digitization noise values and is fit for detecting faint objects. On the other hand, the high gain values are suitable for observing bright objects and evading saturation.

\begin{table*}[t]
\tabularfont
\caption{gain and $RN$ values are verified by generating the PTCs and using the Janesick method and compared with the theoretical ones (provided by the vendor STA). The measured values computed using the PTCs and Janesick methods are consistent with each other and with theoretical ones within a few \%. The present values are also compatible with those reported in \textcolor{blue}{P18}\href{https://ui.adsabs.harvard.edu/abs/2018BSRSL..87...42P/abstract}~for readout speed of 1 MHz and gain = 1 e$^-$/ADU.}
\label{tab:gain_rn}
\addtolength{\tabcolsep}{13pt}
\begin{tabular}{lccccc}
\topline
Readout & Theoretical & Measured & Measured & Theoretical & Measured \\
Speed & Gain & Gain & Gain & RN & RN\\ 
     &  & (PTC) & (Janesick) &  & (Janesick) \\
$ $ & (e$^{-}$/ADU) & (e$^{-}$/ADU) & (e$^{-}$/ADU) & (e$^{-}$) & (e$^{-}$) \\
\hline 
\hline				
100 KHz&1& 0.98$\pm$0.05 &  0.97$\pm$0.04 & 3 & 2.97$\pm$0.06 \\ 
$ $ & 2 & 1.95 $\pm$0.06 &  1.96$\pm$0.07 & 3 & 2.84$\pm$0.09 \\
$ $ & 3 & 2.87 $\pm$0.10 &  2.92$\pm$0.08 & 4 & 3.32$\pm$0.10 \\
$ $ & 5 & 4.85 $\pm$0.12 &  4.90$\pm$0.10 & 4 & 3.97$\pm$0.12 \\
$ $ & 10 & 9.81$\pm$0.15 &  9.92$\pm$0.21 & 4  & 4.22$\pm$0.14 \\
\hline
500 KHz&1& 0.95$\pm$ 0.06 &  0.97$\pm$0.04 & 5 & 4.98$\pm$0.12 \\
$ $ & 2 & 1.92$\pm$0.06 &  1.95$\pm$0.05 & 5 & 4.78$\pm$0.13 \\
$ $ & 3 & 3.15$\pm$0.11 &  3.08$\pm$0.06 & 6 & 6.15$\pm$0.13 \\
$ $ & 5 & 4.81$\pm$0.15 &  4.94$\pm$0.11 & 6 & 6.43$\pm$0.18 \\
$ $ & 10 &9.78$\pm$0.18 &  9.82$\pm$0.16 & 6 & 6.88$\pm$0.21 \\
\hline
1 MHz &1& 0.91$\pm$0.07 &  0.97$\pm$0.09 &  8 & 7.52$\pm$0.20 \\
$ $ & 2 & 2.18$\pm$0.08 &  2.06$\pm$0.05 &  8 & 7.56$\pm$0.21 \\
$ $ & 3 & 3.19$\pm$0.15 &  3.10$\pm$0.05 &  10 & 10.07$\pm$0.25 \\
$ $ & 5 & 4.78$\pm$0.15 &  4.95$\pm$0.08 &  10 & 10.20$\pm$0.24 \\
$ $ & 10 &9.51$\pm$0.20 &  9.84$\pm$0.21 &  10 & 10.26$\pm$0.29 \\
\hline
\end{tabular}
\end{table*}

$RN$ is the noise generated by the amplifier of the CCD chip during the conversion of the stored charge of the individual pixel into an analog voltage used by the ADC to generate ADUs. RN is usually quoted in terms of the number of electrons introduced per pixel in the final signal. RN is independent of signal, Gaussian in nature, and increases with the increasing readout speed.\\

As a part of the present analysis, Photon Transfer Curves (PTCs) and the Janesick method \citep{Janesick2001} were used to estimate and verify the Gain values; however, the $RN$ values were calculated only using the Janesick method using the sky flat/bias frames acquired with the 4K$\times$4K CCD Imager:\\

1) Photon Transfer Curves:\\

PTC depicts a linear relation between the mean counts and the variance of counts from the mean values. Using PTCs, gain values for a CCD can be determined using the formula:

\begin{equation}\label{eq:mag3}
\sigma_{total}^2(ADU) = \sigma_{rn}^2(ADU) + \frac{N}{Gain}
\end{equation}

Here $\sigma_{total}^2$ is the variance of the counts from the mean value, $\sigma_{rn}^2$ is the readout amplifier noise, and $N$ is the mean pixel counts. Equation~\ref{eq:mag3} is a linear relationship between the $N$ and $\sigma_{total}^2$ and, hence, the gain can be determined by the inverse of slope from the linear fit between $N$ and $\sigma_{total}^2$. 

To generate the PTCs, we observed multiple twilight sky flat frames with varying mean counts within linearity regions in 2021. The PTCs were generated using the bias-corrected flat frames at readout speeds of 100 kHz, 500 kHz, and 1 MHz for all the available gain settings (1, 2, 3, 5, and 10 e$^-$/ADU). PTCs obtained for readout speed of 100 kHz and all five gain values are shown in Figure~\ref{fig:gain_ptc} as an example. For each combination, the gain values are estimated with the 1/slope values between the mean counts and variance calculated for smaller and cleaned (free from column defects, hot pixels, non-uniformity, and dead pixels) regions of 100$\times$100 pixels at a minimum of five different locations of the chip. The estimated values of gains thus derived for the 4K$\times$4K CCD Imager are tabulated in Table~\ref{tab:gain_rn}, pretty close to the theoretical gain values (provided by STA, the manufacturer).\\

\begin{table}[t]
\tabularfont
\caption{Theoretical values of the extinction coefficients $\kappa_{ray}(\lambda, h)$, $\kappa_{oz}(\lambda)$, $\kappa_{aer}(\lambda, h)$, and $\kappa_{tot}(\lambda, h)$ (in mag airmass$^{-1}$) estimated using Equations~2, 3, 5, and 6 of \cite{Stalin2008}, respectively for the Devasthal site in Bessell (UBVRI) and SDSS ($ugriz$) filters.}
\addtolength{\tabcolsep}{3pt}
\begin{tabular}{lccccc}
\hline
Filter&$\lambda_{rf} ($\AA$)$&$\kappa_{ray}$&$\kappa_{aer}$&$\kappa_{oz}$&$\kappa_{total}$\\
\hline 
\hline				
  $U$ & 3663 & 0.404 & 0.001 & 0.038 & 0.442 \\
  $B$ & 4361 & 0.198 & 0.001 & 0.033 & 0.232 \\
  $V$ & 5448 & 0.081 & 0.026 & 0.028 & 0.134 \\
  $R$ & 6407 & 0.042 & 0.000 & 0.024 & 0.066 \\
  $I$ & 7980 & 0.017 & 0.000 & 0.020 & 0.038 \\
  \hline
  $u$ & 3596 & 0.435 & 0.002 & 0.038 & 0.476 \\
  $g$ & 4639 & 0.154 & 0.002 & 0.031 & 0.188 \\
  $r$ & 6122 & 0.050 & 0.000 & 0.025 & 0.075 \\
  $i$ & 7439 & 0.023 & 0.000 & 0.022 & 0.044 \\
  $z$ & 8896 & 0.011 & 0.000 & 0.019 & 0.030 \\
\hline
\label{tab:extinction_theoritical}
\end{tabular}
\end{table}

2) Janesick Method:\\

Under the Janesick Method \citep{Janesick2001}, at least two bias and two flat frames are needed to estimate the gain and RN values. Hence, the bias and flat frames were observed, adapting each readout speed and gain combination. First, we selected a clean region of 100 $\times$ 100 pixels for both the flat and bias frames. Then, we estimated mean counts for both the flat ($\overline{F_1}$ and $\overline{F_2}$) and the bias frames ($\overline{B_1}$ and $\overline{B_2}$). The difference between the two flat frames was created by subtracting two flat frames to remove flat field variations and estimated the variance of the difference frame ($\sigma^2_{F_1 - F_2}$) and likewise for bias images ($\sigma^2_{B_1 - B_2}$). The gain and RN of the STA CCD chip were then calculated as:

\begin{equation}\label{eq:mag4}
Gain = \frac{(\overline{F_1} + \overline{F_2}) - (\overline{B_1} + \overline{B_2})}{\sigma^2_{F_1 - F_2} - \sigma^2_{B_1 - B_2}}
\end{equation}

\begin{equation}\label{eq:mag5}
RN = \frac{gain.\sigma_{B_1 - B_2}}{\sqrt{2}}
\end{equation}

The gain values estimated using the Janesick method are consistent with those calculated by generating the PTCs (see Table~\ref{tab:gain_rn}). To evaluate the $RN$ values by adapting the Janesick method using Equation~\ref{eq:mag5}, we took nearly a hundred bias frames for each combination of readout speeds and gains. We generated the difference bias images by subtracting consecutive bias images, and $\sigma_{B_1 - B_2}$ values were estimated by selecting a clean area of 100 $\times$ 100 pixels within the frames. The $RN$ values were then calculated by substituting mean $\sigma_{B_1 - B_2}$ value from all difference bias images to Equation~\ref{eq:mag5}. The same procedure is followed for all readout speeds (100 kHz, 500 kHz, 1 MHz) and gain settings (1, 2, 3, 5, and 10 e$^-$/ADU). The estimated RN values thus obtained are found to be consistent with those of theoretical ones (see Table~\ref{tab:gain_rn}).

\begin{table*}[t]
\tabularfont
\caption{Extinction coefficients in mag airmass$^{-1}$ for Bessell ($UBVRI$) and SDSS ($ugriz$) filters for the 3.6m DOT at 11m from the ground at the Devasthal site.}
\label{tab:extinction_coefficient}
\addtolength{\tabcolsep}{11pt}
\begin{tabular}{lccccc}
\topline
\hline
    $ $  & $u$ & $g$ & $r$ & $i$ & $z$ \\
  2017-04-16 (SA~104) & 0.65$\pm$0.03 & 0.27$\pm$0.03 & 0.19$\pm$0.08 & 0.14$\pm$0.08 & 0.09$\pm$0.10 \\
  2021-01-25 (PG0918) & 0.49$\pm$0.03 & 0.20$\pm$0.01 & 0.09$\pm$0.01 & 0.08$\pm$0.02 & 0.05$\pm$0.02 \\
  \hline
  $ $ & $U$ & $B$ & $V$ & $R$ & $I$ \\
  2017-04-16 (PG0918) & 0.58$\pm$0.06 & 0.36$\pm$0.05 & 0.28$\pm$0.04 & 0.22$\pm$0.03 & 0.16$\pm$0.03 \\
  2018-03-27 (PG~1323) & --$\pm$-- & 0.36$\pm$0.01 & 0.27$\pm$0.03 & 0.22$\pm$0.03 & 0.16$\pm$0.02 \\
  \hline
  2021-01-25 (PG0918) & 0.43$\pm$0.01 & 0.22$\pm$0.01 & 0.14$\pm$0.01 & 0.08$\pm$0.01 & 0.03$\pm$0.02 \\
  2021-02-06 (PG~1657) & 0.44$\pm$0.02 & 0.22$\pm$0.01 & 0.11$\pm$0.02 & 0.07$\pm$0.02 & 0.03$\pm$0.02 \\
  2021-02-07 (PG~1657) & 0.43$\pm$0.01 & 0.21$\pm$0.01 & 0.13$\pm$0.01 & 0.07$\pm$0.02 & 0.03$\pm$0.02 \\
  Mean & 0.43$\pm$0.01 & 0.21$\pm$0.01 & 0.12$\pm$0.01 & 0.07$\pm$0.01 & 0.03$\pm$0.01 \\
\hline
\end{tabular}
\end{table*}

\begin{figure*}[!h]
\includegraphics[height=.487\textheight]{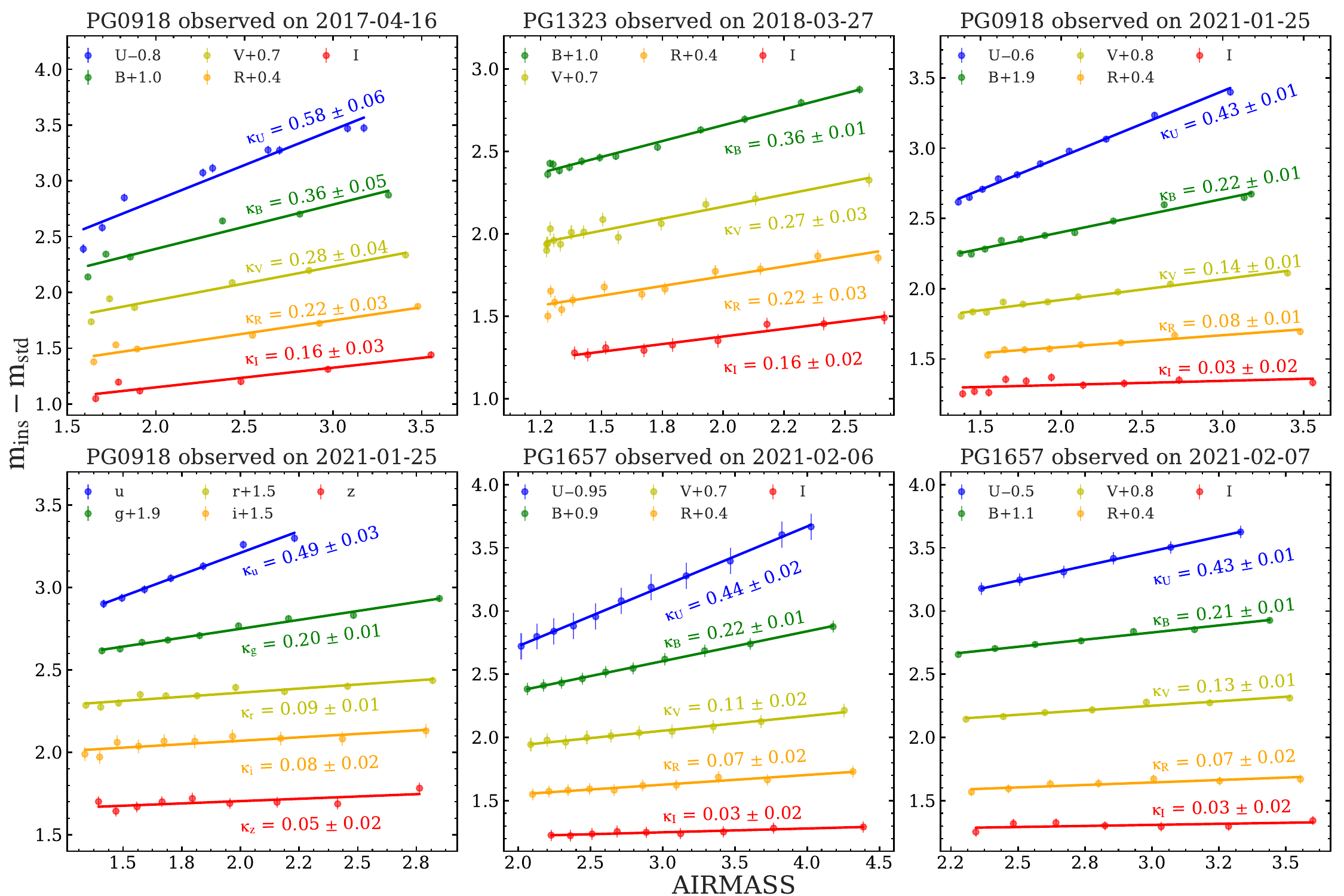}
\caption{The $m(\lambda, z) - m_{o}(\lambda)$ versus airmass values are presented for some of the Landolt fields observed on 2017-04-16, 2018-03-28, 2021-01-25, 2021-02-07, and 2021-02-08. The straight lines represent the linear fits to the $m(\lambda, z) - m_{o}(\lambda)$ versus airmass, that provide the slope values which corresponds to $\kappa_\lambda$ $\times$ 1.086. The extinction coefficients values ($\kappa_\lambda$ = slope/1.086) in Bessell and SDSS filters observed on 2017-04-16, 2018-03-28, 2021-01-25, 2021-02-07, and 2021-02-08 are also mentioned.}
\label{fig:2017_18_extinction}
\label{fig:final_extinction}
\end{figure*}

\begin{figure*}[!h]
\centering
\includegraphics[height=.32\textheight]{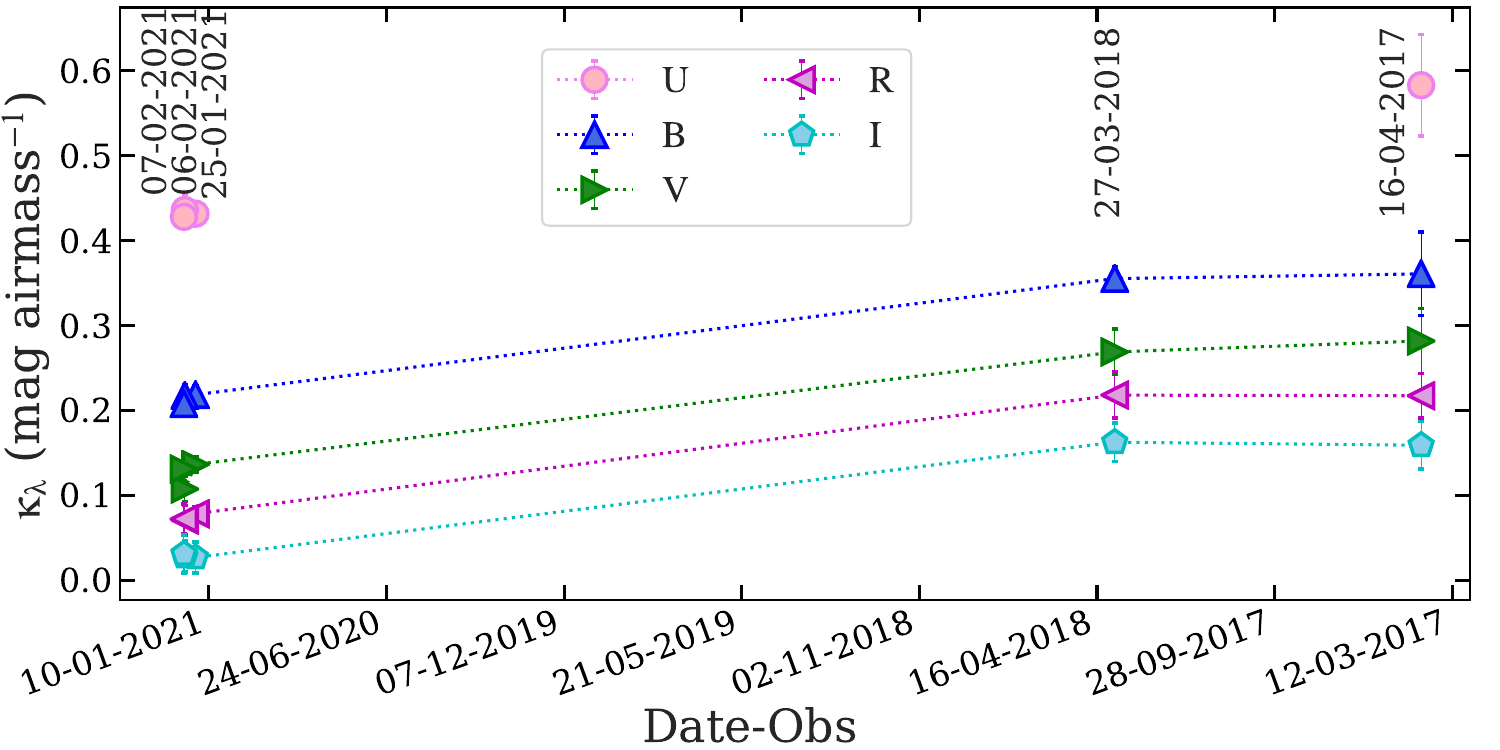}
\caption{Extinction coefficient values in Bessell $UBVRI$ filters for Devasthal site (adopted from Table~\ref{tab:extinction_coefficient}) are plotted as a function of time of observations. The figure indicates the possible seasonal variation of the extinction values in all five filters, i.e., during summer months (March-April), the values are rather higher than those observed during winters (January-February).}
\label{fig:extinction_test}
\end{figure*}

\begin{figure*}[!h]
\centering
\includegraphics[height=.32\textheight]{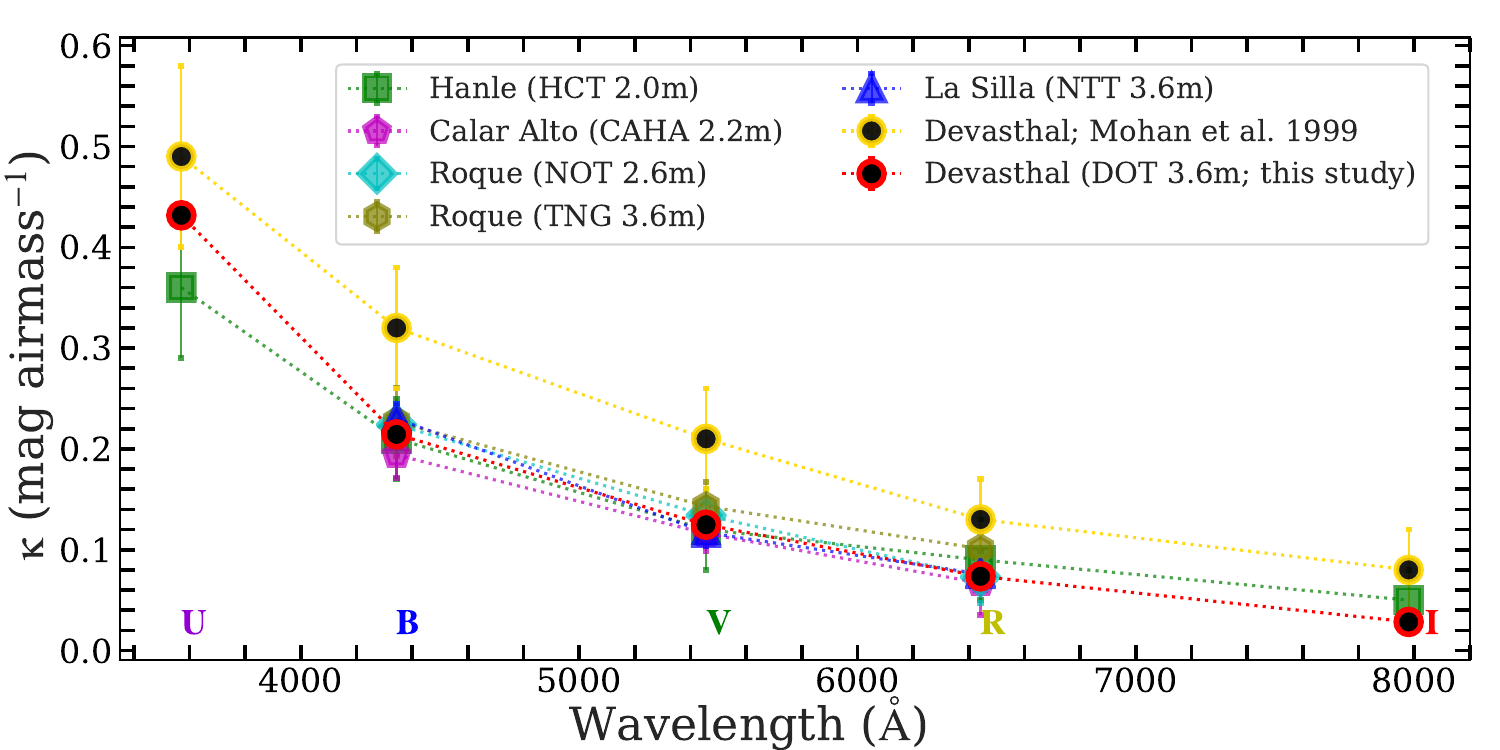}
\caption{Extinction coefficient values of Bessell $UBVRI$ filters for Devasthal site compared with other good astronomical sites hosting 2 to 4-m class telescopes reported by \cite{Stalin2008} and \cite{Altavilla2021}.}
\label{fig:extinction_comp}
\end{figure*}

\section{Atmospheric extinction coefficients}\label{sec:extinction}

The light coming from celestial objects travels through the Earth's atmosphere and is generally absorbed or scattered by the air molecules, aerosols, water vapor, and ozone \citep{Hayes1975}. The amount of attenuation depends on various factors, including the wavelength of incoming light, constituents of the atmosphere, and the site's altitude. However, the atmospheric extinction mainly depends on Rayleigh scattering ($\kappa_{ray}$), molecular absorption by ozone molecules ($\kappa_{oz}$), and aerosol scattering ($\kappa_{aer}$); see \cite{Hayes1975}. All three discussed coefficients are wavelength ($\lambda$) dependent, whereas $\kappa_{ray}$ and $\kappa_{aer}$ are additionally dependent on the altitude of the site ($h$). The values of $\kappa_{ray}(\lambda, h)$, $\kappa_{oz}(\lambda)$, and $\kappa_{aer}(\lambda, h)$ are estimated using Equations~2, 3, and 5 of \cite{Stalin2008}, respectively, by adopting $h$ = 2450 m for the Devasthal site and the values are tabulated in Table~\ref{tab:extinction_theoritical}.

The total extinction coefficient value is a linear combination of the $\kappa_{ray}(\lambda, h)$, $\kappa_{oz}(\lambda)$, and $\kappa_{aer}(\lambda, h)$ and can be expressed as:

\begin{equation}\label{eq:a_A_tot}
\kappa_{tot}(\lambda, h) = \kappa_{ray}(\lambda, h) + \kappa_{oz}(\lambda) + \kappa_{aer}(\lambda, h)
\end{equation}

The theoretical values of total extinction coefficient for Bessell $U$, $B$, $V$, $R$, and $I$ filters are 0.44, 0.23, 0.13, 0.07, and 0.04 mag airmass$^{-1}$, respectively, whereas for SDSS  $u$, $g$, $r$, $i$, and $z$ filters the values are 0.48, 0.19, 0.08, 0.04, and 0.03 mag airmass$^{-1}$, respectively.\\ 

In addition, we also used Bouguer's linear equation to estimate the atmospheric extinction values using photometric observations.

\vspace{0.5cm}

\begin{equation}\label{eq:mag8}
m(\lambda, z) = m_{o}(\lambda) + 1.086 \times \kappa_\lambda \times sec(z)
\end{equation}

Here m($\lambda$, z) is the observed apparent magnitude, $m_o(\lambda)$ is the standard magnitude above Earth's atmosphere, $\kappa_\lambda$ is extinction, $z$ is the zenith angle, and $sec(z)$ represents the airmass.

Equation~\ref{eq:mag8} is used to estimate the extinction coefficient values in all ten broadband filters (Bessell and SDSS) for the 3.6m DOT site (around 11m height from the ground, at the top of the telescope pier). The data used to estimate extinction values thus comprise a total of 6 sets in each of the $UBVRI$/$ugriz$ filters, covering an airmass range of $\sim$1.3 to 4.4. The $m(\lambda, z) - m_{o}(\lambda)$ versus airmass provides the slope values that corresponds to $\kappa_\lambda$ $\times$ 1.086. The $\kappa_\lambda$ (slope/1.086) values estimated using the data observed on 2017-04-16 (PG 0918 for $UBVRI$), 2018-03-27 (PG~1323 for $BVRI$), 2021-01-25 (PG 0918 for $UBVRI$ and $ugriz$), 2021-02-06 (PG~1657 for $UBVRI$), and 2021-02-07 (PG~1657 for $UBVRI$) are shown in Figure~\ref{fig:final_extinction} and the results obtained are tabulated in Table~\ref{tab:extinction_coefficient}. The extinction coefficients determined using the SA~104 field observed on 2017-04-16 in $ugriz$ filters are not shown in Figure~\ref{fig:final_extinction} due to lower data coverage; however the values are listed in Table~\ref{tab:extinction_coefficient}. Using these observations, the observed values of extinction coefficients based on the data obtained in 2021 winters (Bessell; $U$ = 0.43$\pm$0.01, $B$ = 0.21$\pm$0.01, $V$ = 0.12$\pm$0.01, $R$ = 0.07$\pm$0.01, and $I$ = 0.03$\pm$0.01 mag airmass$^{-1}$, SDSS; $u$ = 0.48$\pm$0.03, $g$ = 0.20$\pm$0.01, $r$ = 0.09$\pm$0.01, $i$ = 0.08$\pm$0.02, and $z$ = 0.05$\pm$0.02 mag airmass$^{-1}$) are consistent with those calculated theoretically using Equation~\ref{eq:a_A_tot} (see Table~\ref{tab:extinction_theoritical}). We caution here that, for the SDSS filters, the data available to calculate the extinction coefficients are for only one night (2021-01-25). The extinction coefficients estimated during the winters of 2021 in the $UBVRI$ filters for three different nights were consistent with each other but were found to be lower than those measured in the summer months of 2017 and 2018, as shown in Figure~\ref{fig:extinction_test}. This is consistent with the seasonal variation of the extinction coefficients in different broadband filters and was also reported by \cite{Kumar2000} for Manora peak, one of the other observational sites in the same region of Himalayas. The higher values of extinction coefficients during the summer months might reflect the higher dust content in the atmosphere migrating from the plane area with the prevailing wind and other possible reasons like forest fires, etc.; however, a detailed analysis of these possible reasons is beyond the scope of this paper.

The extinction coefficients measured for the Devasthal site in five Bessell filters presented in this study are also compared with corresponding values from some of the other well-known astronomical sites: Hanle (2.0m Himalayan Chandra Telescope, HCT-2.0m), Calar Alto (2.2m-telescope on Calar Alto; CAHA 2.2m), Roque (2.6m Nordic Optical Telescope; NOT 2.6m and 3.6m Telescopio Nazionale Galileo; TNG 3.6m), and La Silla (3.6m New Technology Telescope; NTT 3.6m). The values of extinction coefficients shown in Figure~\ref{fig:extinction_comp} for the Hanle site are taken from \cite{Stalin2008}, and for all other sites, the values are adopted from \cite{Altavilla2021} and references therein. The extinction coefficients for the Devasthal site are in the range of those found for other good astronomical sites discussed but lower than those reported by \cite{Mohan1999}; see Figure~\ref{fig:extinction_comp}. It is also worth mentioning that the extinction coefficients reported in \cite{Mohan1999} were measured at ground level and a little far away from the 3.6m DOT building, so they can not be compared directly with the present values. The extinction coefficients re-calculated in this study using the data observed in 2017--2018 are consistent with the extinction coefficient values reported by \textcolor{blue}{P18}\href{https://ui.adsabs.harvard.edu/abs/2018BSRSL..87...42P/abstract}.

\vspace{1cm}
\section{Photometric transformation coefficients\label{sec:photo_calib}}

Photometric systems are defined by the filters and detectors used for specific astronomical observations and science goals. To calibrate magnitudes of celestial objects in a standard photometric system, it is required to observe standard stars with a wide range of colors and brightness values \citep[e.g.,][]{Landolt1992} along with science fields. The correction terms estimated using the two sets of values, i.e., standard and observed magnitudes, are called transformation coefficients, and the procedure is labeled as photometric calibration.

We estimate the transformation coefficients assuming a linear relationship between the standard and observed magnitudes that are corrected for the atmospheric extinction, airmass, and exposure times \citep{Romanishin2002}. The five color-color (Equations~\ref{eq:mag9a} to \ref{eq:mag9e}) and five color-mag equations (Equations~\ref{eq:mag9aappendix} to \ref{eq:mag9e_appendix}) given below are used to determine the transformation coefficients.

\begin{subequations}\label{eq:mag9}
\setstretch{-1}
\begin{equation}\label{eq:mag9a}
(B - V)_{std} = \alpha_0 (B - V)_{obs} + \beta_0 \\
\end{equation}
\vspace*{-3mm}
\begin{equation}\label{eq:mag9b}
(U - B)_{std} = \alpha_1 (U - B)_{obs} + \beta_1 \\
\end{equation}
\vspace*{-3mm}
\begin{equation}\label{eq:mag9c}
(V - R)_{std} = \alpha_2 (V - R)_{obs} + \beta_2 \\
\end{equation}
\vspace*{-3mm}
\begin{equation}\label{eq:mag9d}
(R - I)_{std} = \alpha_3 (R - I)_{obs} + \beta_3 \\
\end{equation}
\vspace*{-3mm}
\begin{equation}\label{eq:mag9e}
(V - I)_{std} = \alpha_4 (V - I)_{obs} + \beta_4 \\
\end{equation}
\end{subequations}

\begin{subequations}\label{eq:mag9_appendix}
\setstretch{-1}
\begin{equation}\label{eq:mag9aappendix}
U_{ccd} = U_{std} + \alpha_5 (U - B)_{std} + \beta_5 \\
\end{equation}
\vspace*{-3mm}
\begin{equation}\label{eq:mag9b_appendix}
B_{ccd} = B_{std} + \alpha_6 (B - V)_{std} + \beta_6  \\
\end{equation}
\vspace*{-3mm}
\begin{equation}\label{eq:mag9c_appendix}
V_{ccd} = V_{std} + \alpha_7 (V - R)_{std} + \beta_7 \\
\end{equation}
\vspace*{-3mm}
\begin{equation}\label{eq:mag9d_appendix}
R_{ccd} = R_{std} + \alpha_8 (R - I)_{std} + \beta_8 \\
\end{equation}
\vspace*{-3mm}
\begin{equation}\label{eq:mag9e_appendix}
I_{ccd} = I_{std} + \alpha_9 (V - I)_{std} + \beta_9 \\
\end{equation}
\end{subequations}

\begin{figure}[!h]
\centering
\includegraphics[height=.25\textheight]{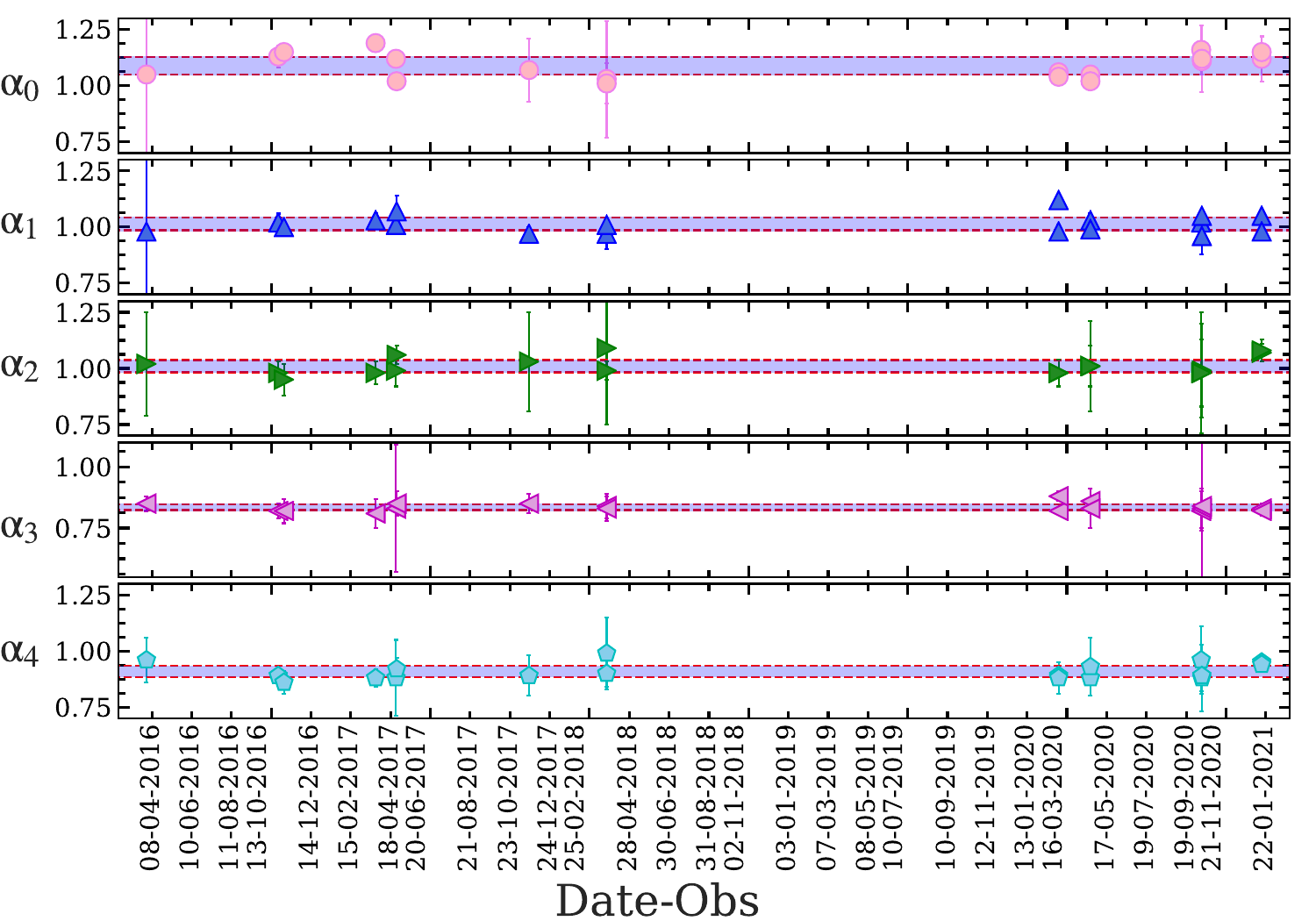}
\caption{Temporal evolution of color coefficients ($\alpha_0$ to $\alpha_4$) as estimated for 18 Landolt fields observed during 13 different good photometric nights in a time span of 5 years (2016 to 2021). The red dotted lines highlight deviations up to 1$\sigma$ (from the mean, close to one). To show the temporal evolution, the data of these selected Landolt fields are taken from Table~\ref{tab:calibration_beta} with observations taken in good photometric sky conditions only (shown in boldface in the very last column of Table~\ref{tab:calibration_beta}).}
\label{fig:alpha_calib}
\end{figure}

\begin{figure}[!h]
\centering
\includegraphics[height=.25\textheight]{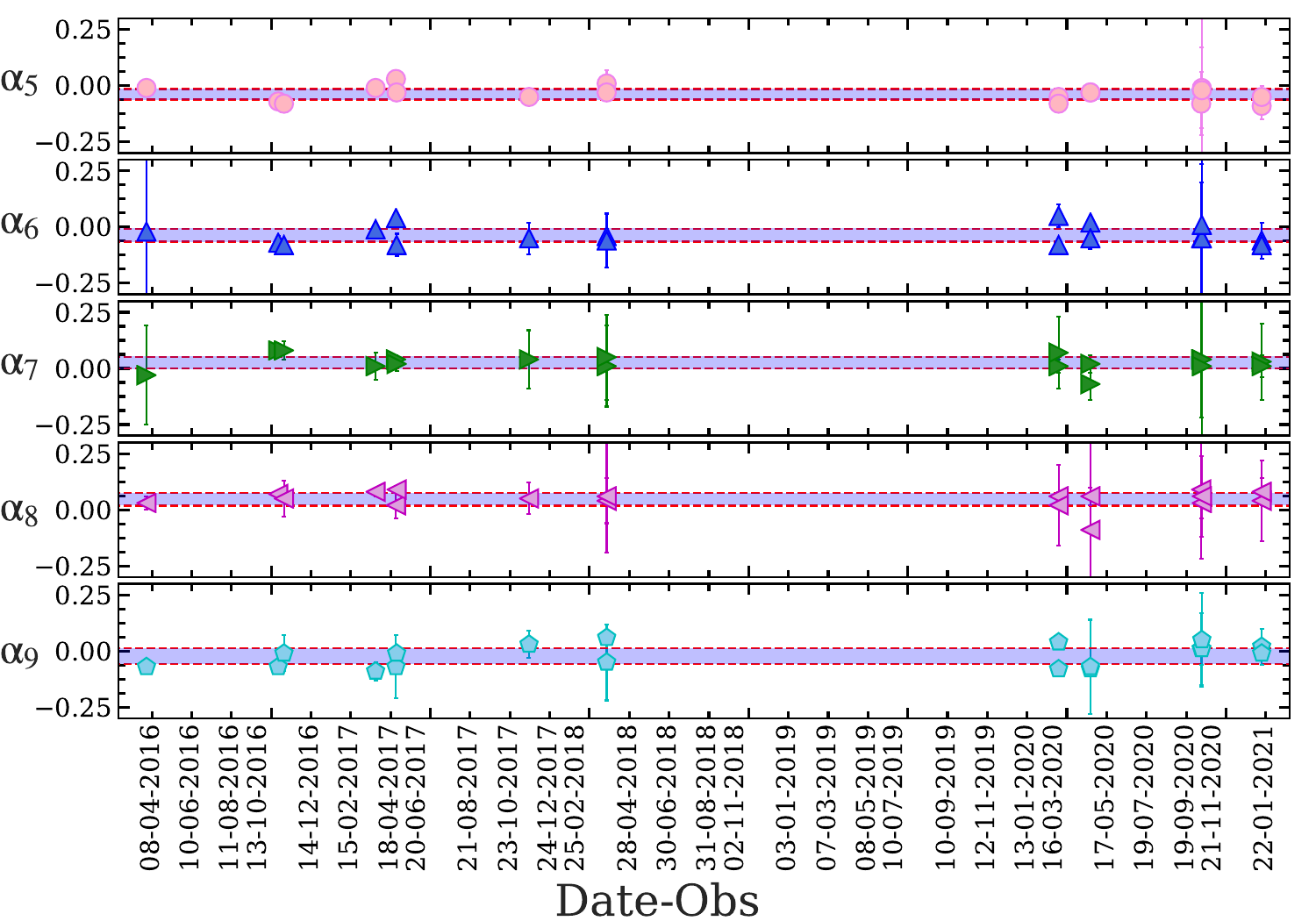}
\caption{Temporal evolution of color coefficients ($\alpha_5$ to $\alpha_{9}$) as estimated for 18 Landolt fields observed during 13 different nights in a time span of 5 years (2016 to 2021). The red dotted lines highlight deviations up to 1$\sigma$ (from the mean, close to zero)}
\label{fig:alpha_calib_appendix}
\end{figure}

\begin{sidewaystable*}
  \centering
  \footnotesize 
  \caption{First-order color coefficients ($\alpha_0$ to $\alpha_4$) and zero-points ($\beta_0$ to $\beta_4$) for color-color relations shown in Equation~\ref{eq:mag9} are calculated for 37 fields distributed over 20 nights from 2016-2021. In the last column, AIRMASS, $M1$ reflectively in percent (Ref.), Moon Phase, and FWHM (in arcsec; for $V$-band) are also quoted to highlight the quality of observing nights.}
  \addtolength{\tabcolsep}{-4.0pt}
  \begin{tabular}{|p{1.15cm}|p{1.55cm}|p{2.85cm}|p{2.85cm}|p{2.85cm}|p{2.85cm}|p{2.95cm}|p{3.15cm}|p{2.95cm}|}
  \hline
  Date&Standard&$\alpha_0$/$\beta_0$&$\alpha_1$/$\beta_1$&$\alpha_2$/$\beta_2$&$\alpha_3$/$\beta_3$&$\alpha_4$/$\beta_4$&AIRMASS/Ref./Moon Phase/FWHM \\
  \hline 
  PG0918&2016-03-31&1.05$\pm$0.71/-1.0$\pm$1.08&0.98$\pm$0.42/-0.63$\pm$0.42&1.02$\pm$0.23/0.07$\pm$0.08&0.85$\pm$0.03/-0.02$\pm$0.02&0.96$\pm$0.1/0.03$\pm$0.08& \bf{1.1208/85/58.86/1.178}   \\ 	 
  PG2213&2016-10-24&1.13$\pm$0.05/0.01$\pm$0.03&1.02$\pm$0.04/-1.7$\pm$0.06&0.98$\pm$0.05/-0.01$\pm$0.02&0.82$\pm$0.03/0.21$\pm$0.01&0.89$\pm$0.03/0.23$\pm$0.02& \bf{1.2131/85/39.90/1.691}    \\ 	 
  PG0231&2016-11-02&1.15$\pm$0.03/-0.01$\pm$0.02&1.0$\pm$0.01/-1.69$\pm$0.02&0.95$\pm$0.07/0.08$\pm$0.04&0.82$\pm$0.05/0.3$\pm$0.02&0.86$\pm$0.05/0.42$\pm$0.05& \bf{1.1214/74.9/8.21/1.406}   \\   
  PG1633&2017-03-26&1.19$\pm$0.03/-0.1$\pm$0.03&1.03$\pm$0.03/-1.83$\pm$0.06&0.98$\pm$0.05/-0.02$\pm$0.03&0.81$\pm$0.06/0.22$\pm$0.02&0.88$\pm$0.04/0.22$\pm$0.03& \bf{1.5125/73.6/2.43/1.254}   \\   
  PG0918&2017-04-16&1.17$\pm$0.02/-0.04$\pm$0.01&1.04$\pm$0.0/-1.8$\pm$0.0&1.0$\pm$0.05/-0.01$\pm$0.02&0.85$\pm$0.04/0.19$\pm$0.01&0.94$\pm$0.01/0.18$\pm$0.01& 1.6608/71.9/74.43/1.406  \\   
  PG1633&2017-04-15&1.05$\pm$0.06/-0.13$\pm$0.05&1.02$\pm$0.07/-1.73$\pm$0.15&1.03$\pm$0.06/-0.13$\pm$0.04&0.75$\pm$0.06/0.21$\pm$0.02&0.88$\pm$0.06/0.13$\pm$0.05& 1.4133/71.9/74.43/1.159 \\   
  SA104 &2017-04-16&1.21$\pm$0.05/-0.1$\pm$0.03&0.93$\pm$0.01/-1.62$\pm$0.02&0.92$\pm$0.06/-0.03$\pm$0.02&0.81$\pm$0.11/0.2$\pm$0.02&0.89$\pm$0.03/0.16$\pm$0.02& 1.1850/71.9/74.43/1.159  \\   
  PG1323&2017-04-27&1.12$\pm$0.0/-0.09$\pm$0.0&1.01$\pm$0.0/-1.73$\pm$0.01&0.99$\pm$0.07/-0.02$\pm$0.02&0.83$\pm$0.26/0.27$\pm$0.07&0.88$\pm$0.17/0.27$\pm$0.08& \bf{1.3005/71.0/7.70/1.368}  \\   
  SA110 &2017-04-28&1.02$\pm$0.03/-0.0$\pm$0.03&1.07$\pm$0.07/-1.97$\pm$0.16&1.06$\pm$0.04/-0.06$\pm$0.03&0.85$\pm$0.05/0.22$\pm$0.04&0.92$\pm$0.05/0.18$\pm$0.07& \bf{1.1837/70.9/15.29/1.368}  \\   
  PG0231&2017-11-21&1.02$\pm$0.13/-0.26$\pm$0.13&0.97$\pm$0.04/-0.98$\pm$0.05&1.05$\pm$0.04/-0.21$\pm$0.03&0.86$\pm$0.02/0.17$\pm$0.01&0.95$\pm$0.02/-0.01$\pm$0.03& 1.7334/56.1/15.73/1.558  \\   
  PG0231&2017-11-22&1.07$\pm$0.14/-0.32$\pm$0.12&0.97$\pm$0.03/-1.45$\pm$0.04&1.03$\pm$0.22/-0.2$\pm$0.16&0.85$\pm$0.04/0.23$\pm$0.02&0.89$\pm$0.09/0.15$\pm$0.1& \bf{1.2145/56.1/23.14/1.121}  \\   
  PG1047&2018-03-24&1.03$\pm$0.26/-1.21$\pm$0.29&0.97$\pm$0.07/-0.92$\pm$0.06&1.09$\pm$0.34/-0.42$\pm$0.2&0.84$\pm$0.05/0.26$\pm$0.01&0.99$\pm$0.16/-0.01$\pm$0.12& \bf{1.1519/49.2/62.73/1.14}  \\   
  PG1525&2018-03-24&1.01$\pm$0.09/-0.63$\pm$0.09&1.01$\pm$0.03/-1.31$\pm$0.04&0.99$\pm$0.04/-0.21$\pm$0.04&0.83$\pm$0.05/0.33$\pm$0.01&0.90$\pm$0.06/0.12$\pm$0.07& \bf{1.2538/49.2/62.73/1.159}  \\   
  PG1323&2020-03-03&1.06$\pm$0.01/0.03$\pm$0.01&0.98$\pm$0.03/-1.58$\pm$0.04&0.98$\pm$0.01/0.04$\pm$0.0&0.82$\pm$0.01/0.25$\pm$0.0&0.89$\pm$0.01/0.31$\pm$0.0& \bf{1.3034/66.7/69.62/1.33}  \\   
  PG1525&2020-03-03&1.04$\pm$0.03/0.06$\pm$0.02&1.12$\pm$0.02/-1.72$\pm$0.02&0.98$\pm$0.06/0.19$\pm$0.02&0.88$\pm$0.02/0.29$\pm$0.0&0.88$\pm$0.07/0.45$\pm$0.05& \bf{1.2474/66.7/69.62/1.083}  \\   
  PG1323&2020-04-22&1.05$\pm$0.04/0.02$\pm$0.02&1.03$\pm$0.03/-1.67$\pm$0.04&1.01$\pm$0.09/0.01$\pm$0.03&0.86$\pm$0.05/0.28$\pm$0.01&0.88$\pm$0.03/0.34$\pm$0.01& \bf{1.3179/65.1/0.51/1.273}   \\   
  STD104&2020-04-22&1.02$\pm$0.04/0.06$\pm$0.02&0.99$\pm$0.01/-1.26$\pm$0.01&1.01$\pm$0.2/0.92$\pm$0.1&0.83$\pm$0.08/0.3$\pm$0.01&0.93$\pm$0.13/1.11$\pm$0.06& \bf{1.1565/65.1/0.51/1.121}   \\   
  STD110&2020-10-03&1.17$\pm$0.02/-0.13$\pm$0.01&0.98$\pm$0.1/-1.63$\pm$0.22&0.99$\pm$0.01/-0.03$\pm$0.0&0.95$\pm$0.12/0.01$\pm$0.06&1.08$\pm$0.04/0.01$\pm$0.05& 1.6054/60.8/96.70/1.786  \\   
  STD111&2020-10-03&1.15$\pm$0.0/-0.04$\pm$0.0&0.95$\pm$0.0/-1.5$\pm$0.0&1.01$\pm$0.0/-0.04$\pm$0.0&0.99$\pm$0.0/0.13$\pm$0.0&1.0$\pm$0.0/0.09$\pm$0.0& 1.2828/60.8/96.70/1.672  \\   
  STD113&2020-10-03&0.93$\pm$0.0/0.14$\pm$0.0&1.28$\pm$0.0/-2.29$\pm$0.0&1.1$\pm$0.0/-0.08$\pm$0.0&0.86$\pm$0.0/0.21$\pm$0.0&0.96$\pm$0.0/0.16$\pm$0.0& 1.1489/60.8/96.70/2.014 \\   
  PG0231&2020-10-04&1.12$\pm$0.03/-0.1$\pm$0.03&1.03$\pm$0.03/-1.66$\pm$0.05&0.97$\pm$0.04/0.03$\pm$0.03&0.87$\pm$0.02/0.21$\pm$0.01&0.96$\pm$0.03/0.19$\pm$0.03& 1.4866/60.8/95.62/1.444  \\   
  STD98 &2020-10-04&1.14$\pm$0.03/-0.01$\pm$0.03&0.94$\pm$0.03/-1.57$\pm$0.07&1.22$\pm$0.16/-0.21$\pm$0.1&1.17$\pm$0.99/-0.16$\pm$0.4&1.23$\pm$0.41/-0.38$\pm$0.44& 1.1791/60.8/92.62/1.387  \\   
  PG1633&2020-10-05&1.15$\pm$0.01/-0.01$\pm$0.01&1.02$\pm$0.01/-1.75$\pm$0.02&0.99$\pm$0.02/0.0$\pm$0.01&0.83$\pm$0.04/0.22$\pm$0.01&0.92$\pm$0.02/0.23$\pm$0.02& 1.6252/60.8/87.01/1.273  \\   
  PG2332&2020-10-05&0.87$\pm$0.0/0.06$\pm$0.0&1.13$\pm$0.14/-1.8$\pm$0.23&1.04$\pm$0.29/-0.08$\pm$0.13&0.79$\pm$0.06/0.23$\pm$0.02&0.9$\pm$0.1/0.2$\pm$0.07& 1.2520/60.8/87.01/1.121 \\   
  STD110&2020-10-05&1.13$\pm$0.03/-0.01$\pm$0.04&0.98$\pm$0.07/-1.59$\pm$0.15&1.05$\pm$0.04/-0.03$\pm$0.03&1.27$\pm$0.33/-0.1$\pm$0.17&1.25$\pm$0.1/-0.11$\pm$0.12& 1.1877/60.8/87.01/1.292 \\   
  PG0231&2020-10-13&1.16$\pm$0.11/-0.06$\pm$0.08&1.02$\pm$0.02/-1.78$\pm$0.04&0.98$\pm$0.27/0.01$\pm$0.19&0.82$\pm$0.08/0.44$\pm$0.03&0.96$\pm$0.15/0.55$\pm$0.12& \bf{1.1297/60.6/13.48/1.14}  \\   
  PG0231&2020-10-14&1.11$\pm$0.04/0.03$\pm$0.03&1.05$\pm$0.02/-1.58$\pm$0.03&0.99$\pm$0.21/0.35$\pm$0.07&0.83$\pm$0.08/0.21$\pm$0.05&0.88$\pm$0.15/0.67$\pm$0.1& \bf{1.1042/60.6/6.47/0.627}   \\   
  STD92 &2020-10-14&1.12$\pm$0.15/0.19$\pm$0.08&0.96$\pm$0.08/-1.73$\pm$0.15&0.98$\pm$0.15/0.23$\pm$0.09&0.84$\pm$0.32/0.35$\pm$0.04&0.89$\pm$0.07/0.48$\pm$0.05& \bf{1.2532/60.6/6.47/0.969}   \\   
  PG0918&2021-01-16&1.12$\pm$0.1/-0.04$\pm$0.05&1.05$\pm$0.01/-1.75$\pm$0.02&1.07$\pm$0.04/0.01$\pm$0.01&0.83$\pm$0.02/0.21$\pm$0.01&0.95$\pm$0.03/0.24$\pm$0.02& \bf{1.1359/59.0/1.49/1.159}   \\   
  PG1323&2021-01-16&1.15$\pm$0.02/-0.01$\pm$0.01&0.98$\pm$0.02/-1.63$\pm$0.03&1.08$\pm$0.05/-0.02$\pm$0.01&0.82$\pm$0.02/0.39$\pm$0.0&0.94$\pm$0.03/0.43$\pm$0.01& \bf{1.4311/59.0/1.49/1.102}   \\   
  PG0918&2021-01-24&1.24$\pm$0.05/-0.09$\pm$0.03&1.06$\pm$0.02/-1.78$\pm$0.04&1.1$\pm$0.14/0.18$\pm$0.04&0.72$\pm$0.1/0.23$\pm$0.03&0.89$\pm$0.02/0.41$\pm$0.01& 1.1266/58.9/90.33/1.178  \\   
  PG0918&2021-01-25&1.24$\pm$0.03/0.85$\pm$0.01&1.04$\pm$0.02/-2.7$\pm$0.05&0.95$\pm$0.03/0.15$\pm$0.01&0.87$\pm$0.01/0.15$\pm$0.0&0.91$\pm$0.01/0.3$\pm$0.01& 3.1322/58.9/94.42/1.976 \\   
  PG0918&2021-01-27&1.31$\pm$0.03/0.7$\pm$0.01&1.03$\pm$0.03/-2.69$\pm$0.09&0.94$\pm$0.01/0.1$\pm$0.0&0.87$\pm$0.0/0.16$\pm$0.0&0.9$\pm$0.01/0.27$\pm$0.0& 3.1355/58.8/99.88/1.463  \\   
  PG1657&2021-02-06&1.17$\pm$0.01/-0.03$\pm$0.01&1.06$\pm$0.03/-1.78$\pm$0.05&1.01$\pm$0.05/-0.01$\pm$0.02&0.5$\pm$0.0/-0.1$\pm$0.0&0.88$\pm$0.03/-0.48$\pm$0.04& 1.3762/58.7/27.77/2.28 \\   
  PG0918&2021-02-07&1.21$\pm$0.03/-0.03$\pm$0.02&1.05$\pm$0.01/-1.79$\pm$0.02&0.94$\pm$0.05/-0.0$\pm$0.02&0.92$\pm$0.05/0.21$\pm$0.01&0.92$\pm$0.03/0.22$\pm$0.02& 1.8123/58.7/18.32/0.703  \\   
  PG1633&2021-02-07&1.22$\pm$0.01/-0.05$\pm$0.01&1.01$\pm$0.02/-1.77$\pm$0.04&1.0$\pm$0.03/0.01$\pm$0.01&0.85$\pm$0.01/0.19$\pm$0.0&0.93$\pm$0.01/0.22$\pm$0.01& 2.7349/58.7/18.32/1.311  \\   
  PG1657&2021-02-07&1.27$\pm$0.01/-0.09$\pm$0.01&1.05$\pm$0.02/-1.84$\pm$0.03&0.99$\pm$0.1/0.03$\pm$0.04&0.9$\pm$0.0/0.2$\pm$0.0&0.91$\pm$0.06/0.25$\pm$0.04& 3.4394/58.7/18.32/1.71  \\                
  \hline
  \end{tabular}
  \label{tab:calibration_beta}
\end{sidewaystable*}

\begin{sidewaystable*}
  \centering
  \footnotesize
  \caption{First-order color coefficients ($\alpha_5$ to $\alpha_{9}$) and zero-points ($\beta_5$ to $\beta_{9}$) for color-mag relations shown in Equation~\ref{eq:mag9_appendix} are calculated for 37 fields distributed over 20 nights from 2016-2021.}
  \addtolength{\tabcolsep}{-4pt}
  \begin{tabular}{|p{1.15cm}|p{1.55cm}|p{2.85cm}|p{2.85cm}|p{2.85cm}|p{2.85cm}|p{2.95cm}|p{3.15cm}|p{2.95cm}|}
  \hline
  Date & Standard & $\alpha_5$/$\beta_5$ & $\alpha_6$/$\beta_6$ & $\alpha_7$/$\beta_7$ & $\alpha_{8}$/$\beta_{8}$ & $\alpha_{9}$/$\beta_{9}$ & AIRMASS/Ref./Moon Phase/FWHM \\
  \hline 
  PG0918 & 2016-03-31 & -0.01$\pm$0.01/4.21$\pm$0.01 & -0.02$\pm$0.73/3.57$\pm$0.51 & -0.03$\pm$0.22/2.58$\pm$0.07 & 0.03$\pm$0.03/2.63$\pm$0.01 & -0.07$\pm$0.0/2.61$\pm$0.0  &  \bf{1.1208/--/58.86/1.178}   \\ 	 
  PG2213 & 2016-10-24 & -0.07$\pm$0.01/3.83$\pm$0.01 & -0.07$\pm$0.03/2.2$\pm$0.02 & 0.08$\pm$0.03/2.2$\pm$0.01 & 0.07$\pm$0.01/2.2$\pm$0.0 & -0.07$\pm$0.02/2.46$\pm$0.02  &  \bf{1.2131/--/39.90/1.691}    \\ 	 
  PG0231 & 2016-11-02 & -0.08$\pm$0.02/3.14$\pm$0.01 & -0.08$\pm$0.02/1.49$\pm$0.02 & 0.08$\pm$0.04/1.47$\pm$0.02 & 0.05$\pm$0.08/1.54$\pm$0.05 & -0.01$\pm$0.08/1.95$\pm$0.08  &  \bf{1.1214/74.9/8.21/1.406}   \\   
  PG1633 & 2017-03-26 & -0.01$\pm$0.02/2.55$\pm$0.01 & -0.01$\pm$0.01/0.83$\pm$0.01 & 0.01$\pm$0.06/0.75$\pm$0.03 & 0.08$\pm$0.02/0.73$\pm$0.01 & -0.09$\pm$0.04/1.0$\pm$0.04  &  \bf{1.5125/73.6/2.43/1.254}   \\   
  PG0918 & 2017-04-16 & -0.12$\pm$0.01/2.55$\pm$0.0 & -0.12$\pm$0.01/0.87$\pm$0.01 & 0.05$\pm$0.02/0.83$\pm$0.01 & 0.05$\pm$0.04/0.82$\pm$0.01 & -0.05$\pm$0.01/1.03$\pm$0.01  &  1.6608/71.9/74.43/1.406  \\   
  PG1633 & 2017-04-15 & -0.01$\pm$0.06/2.97$\pm$0.05 & 0.01$\pm$0.06/1.29$\pm$0.05 & 0.1$\pm$0.15/1.15$\pm$0.07 & 0.11$\pm$0.15/1.02$\pm$0.06 & -0.09$\pm$0.1/1.3$\pm$0.09  &  1.4133/71.9/74.43/1.159 \\   
  SA104  & 2017-04-16 & 0.01$\pm$0.01/2.56$\pm$0.0 & -0.12$\pm$0.02/0.89$\pm$0.01 & 0.1$\pm$0.1/0.79$\pm$0.04 & 0.05$\pm$0.04/0.74$\pm$0.01 & -0.04$\pm$0.08/0.95$\pm$0.06  &  1.1850/71.9/74.43/1.159  \\   
  PG1323 & 2017-04-27 & 0.03$\pm$0.02/2.06$\pm$0.01 & 0.04$\pm$0.02/0.32$\pm$0.01 & 0.04$\pm$0.02/0.24$\pm$0.01 & 0.02$\pm$0.06/0.22$\pm$0.02 & -0.07$\pm$0.14/0.54$\pm$0.08  &  \bf{1.3005/71.0/7.70/1.368}  \\   
  SA110  & 2017-04-28 & -0.03$\pm$0.03/2.87$\pm$0.02 & -0.08$\pm$0.05/1.08$\pm$0.05 & 0.02$\pm$0.03/1.1$\pm$0.03 & 0.09$\pm$0.01/1.04$\pm$0.01 & -0.01$\pm$0.03/1.27$\pm$0.06  &  \bf{1.1837/70.9/15.29/1.368}  \\   
  PG0231 & 2017-11-21 & -0.1$\pm$0.02/2.23$\pm$0.01 & -0.11$\pm$0.1/1.28$\pm$0.08 & 0.04$\pm$0.03/1.02$\pm$0.02 & 0.07$\pm$0.01/0.82$\pm$0.01 & -0.04$\pm$0.01/1.02$\pm$0.01  &  1.7334/56.1/15.73/1.558  \\   
  PG0231 & 2017-11-22 & -0.05$\pm$0.02/2.7$\pm$0.01 & -0.05$\pm$0.07/1.25$\pm$0.06 & 0.04$\pm$0.13/0.99$\pm$0.08 & 0.05$\pm$0.07/0.75$\pm$0.04 & 0.03$\pm$0.06/1.06$\pm$0.07  &  \bf{1.2145/56.1/23.14/1.121}  \\   
  PG1047 & 2018-03-24 & 0.01$\pm$0.06/3.26$\pm$0.05 & -0.04$\pm$0.03/2.32$\pm$0.02 & 0.01$\pm$0.18/1.42$\pm$0.06 & 0.04$\pm$0.1/1.1$\pm$0.03 & 0.06$\pm$0.03/1.42$\pm$0.02  &  \bf{1.1519/49.2/62.73/1.14}  \\   
  PG1525 & 2018-03-24 & -0.03$\pm$0.03/3.17$\pm$0.03 & -0.06$\pm$0.12/1.88$\pm$0.09 & 0.05$\pm$0.19/1.44$\pm$0.09 & 0.06$\pm$0.25/1.14$\pm$0.11 & -0.05$\pm$0.17/1.57$\pm$0.17  &  \bf{1.2538/49.2/62.73/1.159}  \\   
  PG1323 & 2020-03-03 & -0.05$\pm$0.01/3.44$\pm$0.0 & -0.08$\pm$0.01/1.86$\pm$0.01 & 0.01$\pm$0.03/1.88$\pm$0.01 & 0.06$\pm$0.02/1.93$\pm$0.01 & -0.08$\pm$0.02/2.23$\pm$0.01  &  \bf{1.3034/66.7/69.62/1.33}  \\   
  PG1525 & 2020-03-03 & -0.08$\pm$0.01/3.02$\pm$0.01 & 0.05$\pm$0.05/1.46$\pm$0.04 & 0.07$\pm$0.16/1.48$\pm$0.07 & 0.02$\pm$0.18/1.71$\pm$0.08 & 0.04$\pm$0.03/2.02$\pm$0.03  &  \bf{1.2474/66.7/69.62/1.083}  \\   
  PG1323 & 2020-04-22 & -0.03$\pm$0.01/2.54$\pm$0.01 & 0.02$\pm$0.01/0.93$\pm$0.01 & 0.02$\pm$0.04/0.95$\pm$0.01 & 0.06$\pm$0.04/0.95$\pm$0.01 & -0.08$\pm$0.03/1.33$\pm$0.02  &  \bf{1.3179/65.1/0.51/1.273}   \\   
  STD104 & 2020-04-22 & -0.03$\pm$0.01/3.27$\pm$0.0 & -0.05$\pm$0.05/1.99$\pm$0.03 & -0.07$\pm$0.07/2.13$\pm$0.03 & -0.09$\pm$0.58/3.22$\pm$0.2 & -0.07$\pm$0.21/3.41$\pm$0.15  &  \bf{1.1565/65.1/0.51/1.121}   \\   
  STD110 & 2020-10-03 & -0.1$\pm$0.09/2.89$\pm$0.06 & -0.11$\pm$0.01/1.26$\pm$0.01 & 0.07$\pm$0.03/1.15$\pm$0.02 & 0.06$\pm$0.03/1.12$\pm$0.02 & 0.11$\pm$0.05/1.15$\pm$0.07  &  1.6054/60.8/96.70/1.786  \\   
  STD111 & 2020-10-03 & -0.03$\pm$0.0/2.74$\pm$0.0 & -0.11$\pm$0.0/1.17$\pm$0.0 & 0.03$\pm$0.0/1.13$\pm$0.0 & 0.05$\pm$0.0/1.09$\pm$0.0 & 0.02$\pm$0.0/1.22$\pm$0.0  &  1.2828/60.8/96.70/1.672  \\   
  STD113 & 2020-10-03 & -0.17$\pm$0.0/2.8$\pm$0.0 & 0.11$\pm$0.0/0.93$\pm$0.0 & 0.03$\pm$0.0/1.1$\pm$0.0 & 0.15$\pm$0.0/1.01$\pm$0.0 & -0.02$\pm$0.0/1.27$\pm$0.0  &  1.1489/60.8/96.70/2.014 \\   
  PG0231 & 2020-10-04 & -0.1$\pm$0.01/2.65$\pm$0.01 & -0.11$\pm$0.04/1.09$\pm$0.04 & -0.0$\pm$0.04/1.0$\pm$0.02 & 0.03$\pm$0.05/0.98$\pm$0.03 & -0.05$\pm$0.01/1.22$\pm$0.01  &  1.4866/60.8/95.62/1.444  \\   
  STD98  & 2020-10-04 & -0.01$\pm$0.03/2.58$\pm$0.02 & -0.09$\pm$0.03/0.99$\pm$0.04 & 0.05$\pm$0.06/0.98$\pm$0.04 & 0.23$\pm$0.08/0.79$\pm$0.06 & 0.28$\pm$0.11/0.62$\pm$0.18  &  1.1791/60.8/92.62/1.387  \\   
  PG1633 & 2020-10-05 & -0.09$\pm$0.02/2.64$\pm$0.02 & -0.11$\pm$0.02/0.98$\pm$0.01 & 0.04$\pm$0.05/0.97$\pm$0.02 & 0.03$\pm$0.04/0.98$\pm$0.02 & -0.07$\pm$0.01/1.23$\pm$0.01  &  1.6252/60.8/87.01/1.273  \\   
  PG2332 & 2020-10-05 & -0.06$\pm$0.31/2.76$\pm$0.11 & 0.08$\pm$0.14/1.11$\pm$0.1 & -0.13$\pm$0.21/1.18$\pm$0.08 & -0.09$\pm$0.57/1.11$\pm$0.22 & -0.22$\pm$0.57/1.57$\pm$0.48  &  1.2520/60.8/87.01/1.121 \\   
  STD110 & 2020-10-05 & -0.06$\pm$0.02/2.6$\pm$0.02 & -0.1$\pm$0.05/1.03$\pm$0.05 & 0.0$\pm$0.05/1.03$\pm$0.04 & 0.08$\pm$0.07/0.99$\pm$0.05 & 0.21$\pm$0.07/0.94$\pm$0.11  &  1.1877/60.8/87.01/1.292 \\   
  PG0231 & 2020-10-13 & -0.08$\pm$0.14/3.02$\pm$0.09 & -0.05$\pm$0.25/1.33$\pm$0.2 & 0.04$\pm$0.26/1.08$\pm$0.14 & 0.09$\pm$0.31/0.94$\pm$0.18 & 0.01$\pm$0.16/1.59$\pm$0.19  &  \bf{1.1297/60.6/13.48/1.14}  \\   
  PG0231 & 2020-10-14 & -0.01$\pm$0.18/2.75$\pm$0.09 & -0.05$\pm$0.33/1.27$\pm$0.06 & 0.04$\pm$0.81/0.92$\pm$0.39 & 0.03$\pm$0.07/1.58$\pm$0.04 & 0.01$\pm$0.07/1.75$\pm$0.08  &  \bf{1.1042/60.6/6.47/0.627}   \\   
  STD92  & 2020-10-14 & -0.02$\pm$0.35/2.94$\pm$0.03 & 0.01$\pm$0.45/1.08$\pm$0.38 & 0.01$\pm$0.63/0.91$\pm$0.35 & 0.06$\pm$0.18/1.27$\pm$0.09 & 0.05$\pm$0.21/1.58$\pm$0.22  &  \bf{1.2532/60.6/6.47/0.969}   \\   
  PG0918 & 2021-01-16 & -0.09$\pm$0.06/3.48$\pm$0.05 & -0.06$\pm$0.08/1.84$\pm$0.05 & 0.03$\pm$0.17/1.79$\pm$0.07 & 0.04$\pm$0.18/1.81$\pm$0.07 & 0.02$\pm$0.08/2.05$\pm$0.07  &  \bf{1.1359/59.0/1.49/1.159}   \\   
  PG1323 & 2021-01-16 & -0.05$\pm$0.05/3.43$\pm$0.03 & -0.08$\pm$0.04/1.8$\pm$0.02 & 0.01$\pm$0.05/1.78$\pm$0.02 & 0.08$\pm$0.06/1.77$\pm$0.02 & -0.01$\pm$0.05/2.24$\pm$0.04  &  \bf{1.4311/59.0/1.49/1.102}   \\   
  PG0918 & 2021-01-24 & -0.21$\pm$0.04/2.83$\pm$0.03 & -0.21$\pm$0.07/1.26$\pm$0.05 & -0.01$\pm$0.08/1.18$\pm$0.03 & 0.06$\pm$0.19/1.35$\pm$0.07 & -0.1$\pm$0.03/1.63$\pm$0.03  &  1.1266/58.9/90.33/1.178  \\   
  PG0918 & 2021-01-25 & -0.15$\pm$0.02/2.72$\pm$0.02 & -0.15$\pm$0.01/0.21$\pm$0.01 & 0.07$\pm$0.04/0.89$\pm$0.02 & 0.01$\pm$0.02/1.05$\pm$0.01 & -0.06$\pm$0.01/1.22$\pm$0.01  &  3.1322/58.9/94.42/1.976 \\   
  PG0918 & 2021-01-27 & -0.17$\pm$0.03/3.06$\pm$0.02 & -0.21$\pm$0.02/0.56$\pm$0.01 & 0.05$\pm$0.03/1.1$\pm$0.01 & -0.01$\pm$0.02/1.21$\pm$0.01 & -0.07$\pm$0.02/1.38$\pm$0.01  &  3.1355/58.8/99.88/1.463  \\   
  PG1657 & 2021-02-06 & -0.12$\pm$0.02/3.43$\pm$0.01 & -0.1$\pm$0.02/1.81$\pm$0.01 & 0.11$\pm$0.02/1.78$\pm$0.01 & 0.12$\pm$0.02/1.79$\pm$0.01 & -0.08$\pm$0.04/1.23$\pm$0.03  &  1.3762/58.7/27.77/2.28 \\   
  PG0918 & 2021-02-07 & -0.13$\pm$0.01/2.72$\pm$0.01 & -0.12$\pm$0.01/1.06$\pm$0.01 & 0.1$\pm$0.06/1.04$\pm$0.02 & 0.03$\pm$0.03/1.04$\pm$0.01 & -0.03$\pm$0.01/1.27$\pm$0.01  &  1.8123/58.7/18.32/0.703  \\   
  PG1633 & 2021-02-07 & -0.11$\pm$0.02/2.76$\pm$0.02 & -0.16$\pm$0.02/1.08$\pm$0.02 & 0.05$\pm$0.05/1.04$\pm$0.02 & 0.07$\pm$0.04/1.04$\pm$0.02 & -0.03$\pm$0.02/1.25$\pm$0.02 &  2.7349/58.7/18.32/1.311  \\   
  PG1657 & 2021-02-07 & -0.16$\pm$0.02/2.76$\pm$0.01 & -0.15$\pm$0.02/1.09$\pm$0.01 & 0.12$\pm$0.04/1.01$\pm$0.02 & 0.14$\pm$0.04/1.00$\pm$0.02 & -0.03$\pm$0.1/1.28$\pm$0.09 &  3.4394/58.7/18.32/1.71  \\                
  \hline
  \end{tabular}
  \label{tab:calibration_color_mag}
\end{sidewaystable*}

Magnitudes terms with subscript $obs$ represent the extinction and airmass corrected observed magnitudes, and terms with subscript $std$ stand for standard magnitudes. The quantities $\alpha_0$ -- $\alpha_9$ and $\beta_0$ -- $\beta_9$ are the color coefficients and the zero-points, respectively.

To estimate the $\alpha$ and $\beta$ values as described above, 37 Landolt standard fields \citep{Landolt1992} distributed over 20 different nights were observed in all Bessell filters ($UBVRI$; between 2016 March and 2021 February); see Tables~\ref{tab:calibration_beta} (for color-color relations) and \ref{tab:calibration_color_mag} (for color-mag relations). Eqs. from \ref{eq:mag9a} to \ref{eq:mag9e} and \ref{eq:mag9aappendix} to \ref{eq:mag9e_appendix} are linear relations between color-color and color-mag, respectively. Whereas, the values of $\alpha$'s and $\beta$'s in these linear relations are computed as slopes and intercepts, respectively.

In the present analysis and based on the available data observed at different values of $M1$ reflectively ($\sim$49 -- 85\%), good photometric conditions are considered based on the criteria: clear sky with moon phase $<$ 70\%, humidity $<$ 70\%, AIRMASS $\lesssim$1.5, and FWHM $<$1.7 arcsec ($V$-band; see Figure~\ref{fig:fwhm_three_cycle}). 
The data taken during above-mentioned criteria of photometric night conditions consists 18 Landolt fields distributed over 13 different nights, shown in bold in the last columns of Tables~\ref{tab:calibration_beta} and \ref{tab:calibration_color_mag}. The $\alpha$ and $\beta$ values calculated for a total of 18 such fields are plotted in Figures~\ref{fig:alpha_calib} (for color-color relations) and \ref{fig:alpha_calib_appendix} (for color-mag relations). Standard fields observed on different nights have consistent values of $\alpha$ for a given colour; however, there is a significant scatter in $\beta$ values. It could be because $\alpha$ values depend mainly on the instrument response, whereas $\beta$ values rely on different parameters responsible for observing conditions.

Based on the present analysis, mean values of $\alpha_0$ to $\alpha_9$ plotted in Figures~\ref{fig:alpha_calib} and \ref{fig:alpha_calib_appendix} are: $\alpha_0$ = 1.09$\pm$0.04, $\alpha_1$ = 1.01$\pm$0.03, $\alpha_2$ = 1.01$\pm$0.03, $\alpha_3$ = 0.84$\pm$0.01, $\alpha_4$ = 0.91$\pm$0.02, $\alpha_5$ = $-$0.04$\pm$0.02, $\alpha_6$ = $-$0.04$\pm$0.03, $\alpha_7$ = 0.02$\pm$0.02, $\alpha_8$ = 0.05$\pm$0.03, and $\alpha_9$ = $-$0.02$\pm$0.03. From 2016 to 2021, the values of $\alpha_0$--$\alpha_4$ and $\alpha_5$--$\alpha_9$ seem consistent and within 1$\sigma$ from the mean (shown with red dotted lines in Figures~\ref{fig:alpha_calib} and \ref{fig:alpha_calib_appendix}) respectively to one and zero and do not appear to have any noticeable evolution with time.

\begin{figure*}[!h]
\centering
\includegraphics[height=.35\textheight]{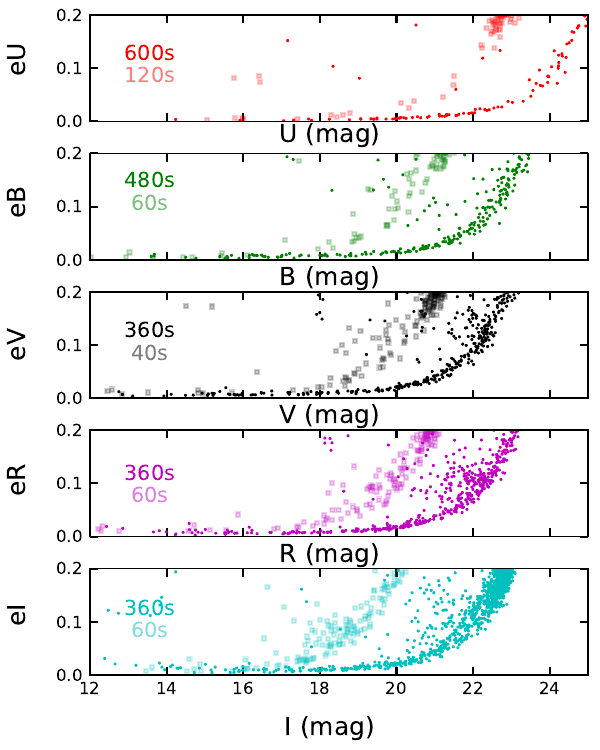}
\includegraphics[height=.35\textheight]{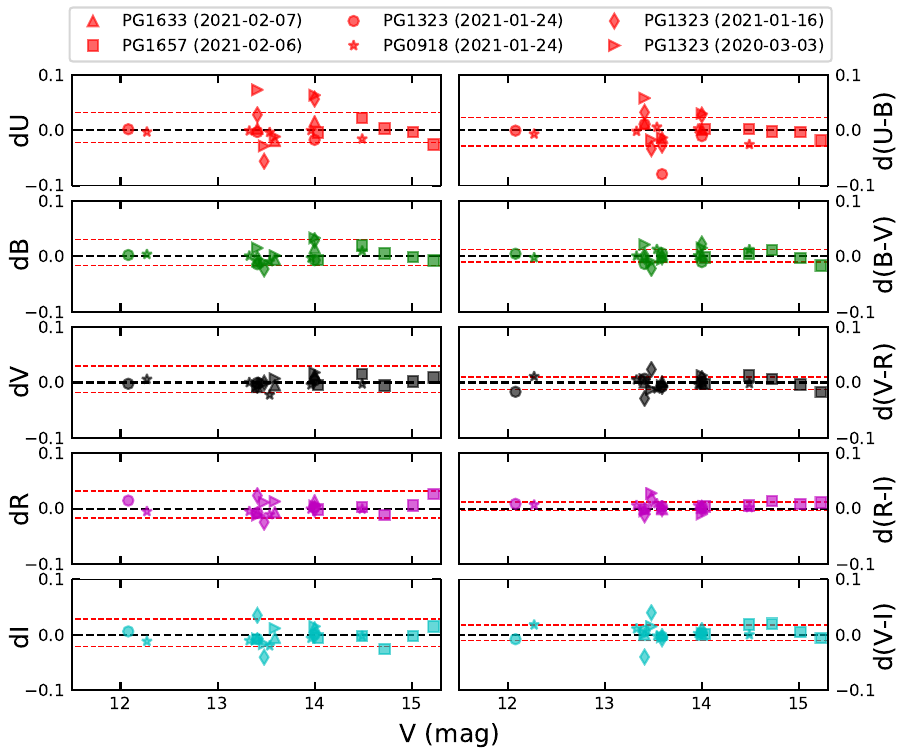}
\caption{Left panel: the field photometry of PG~0231 and SN~2020ank \citep{Kumar2021} for exposures 40--120s and 360--600s, respectively, showing deeper detection for longer exposures as simulated for the Imager (\textcolor{blue}{P18}\href{https://ui.adsabs.harvard.edu/abs/2018BSRSL..87...42P/abstract}). Middle panel: difference of the photometry of Landolt fields PG~1323, PG~0918, PG~1657, and PG~1633 \citep{Landolt1992}, in $UBVRI$ filters. Right panel: difference of the colors of Landolt fields discussed-above, showing a mean value of $\approx$0.01 $\pm$ 0.02 mag for calibrations between the brightness range of $\sim$12 to 15.3 mag.}
\label{fig:field_SLSN2020ank}
\end{figure*}

\subsection{Photometric errors and comparison}
\label{sec:photo_preci}

In the left panel of Figure~\ref{fig:field_SLSN2020ank}, calibrated magnitudes of the stars in the field of PG~0231 \citep[faded squares;][]{Landolt1992} and SN~2020ank \citep[dots;][]{Kumar2021} along with associated photometric errors are plotted for exposures between 60--120s and 360--600s, respectively, for all the filters. The $UBVRI$ bands data of PG~0231 and field of SN~2020ank presented here were observed on 2020 October 14, under $M1$ reflectively of $\sim$60.1\% and moon illumination of $\approx$6.5\%. Figure~\ref{fig:field_SLSN2020ank} indicates that associated photometric errors are $\sim$0.05 mag in $V$ $\sim$19.5 mag for PG0231 (for an exposure time of 40s; FWHM in $V$-band $\sim$0.6 arcsec) whereas it is $\sim$0.05 mag for $V$ $\sim$21.5 mag for SN~2020ank field (for an exposure time of 360s; FWHM in $V$-band $\sim$0.8 arcsec). A similar trend is also followed in other filters, and these magnitude values are close to those of the simulated ones presented in \textcolor{blue}{P18}\href{https://ui.adsabs.harvard.edu/abs/2018BSRSL..87...42P/abstract}. Field photometry of SN~2012au was also performed using the Landolt standard field PG~1323 on 2020 March 03 using the 4K$\times$4K CCD Imager; \cite{Pandey2021} further demonstrated consistent results of photometric errors and limiting magnitudes for given exposure times. Late-time optical observations of GRB~200412B field were also performed by \citet{Kumar2020} based on the data taken using the 4K$\times$4K CCD Imager, and detection was reported at 25.21$\pm$0.10 mag in $g$-band (360s$\times$10 frames) and 24.62 $\pm$ 0.12 mag in $R$-band (360s$\times$12 frames).

To compare the calibrated photometric magnitudes of standard stars with those published for Landolt fields PG~1323, PG~0918, PG~1657, and PG~1633 \citep{Landolt1992}, we plot the difference between the calibrated and standard magnitudes and the difference between their respective colors in the middle and right panels of Figure~\ref{fig:field_SLSN2020ank}. The Landolt fields used in the present analysis exhibit a brightness range of $V$ $\sim$12--15.3 mag distributed over five good photometric nights during 2020--2021. The mean values of the difference in magnitudes and colors for the four Landolt fields mentioned above are $\approx$ 0.01 $\pm$ 0.02 mag. The results show that our photometry is comparable within 1$\sigma$ to those published for Landolt fields in all the filters for the given brightness range (see middle and right panels of Figure~\ref{fig:field_SLSN2020ank}).

\begin{figure*}[!h]
\centering
\includegraphics[height=.7\textheight]{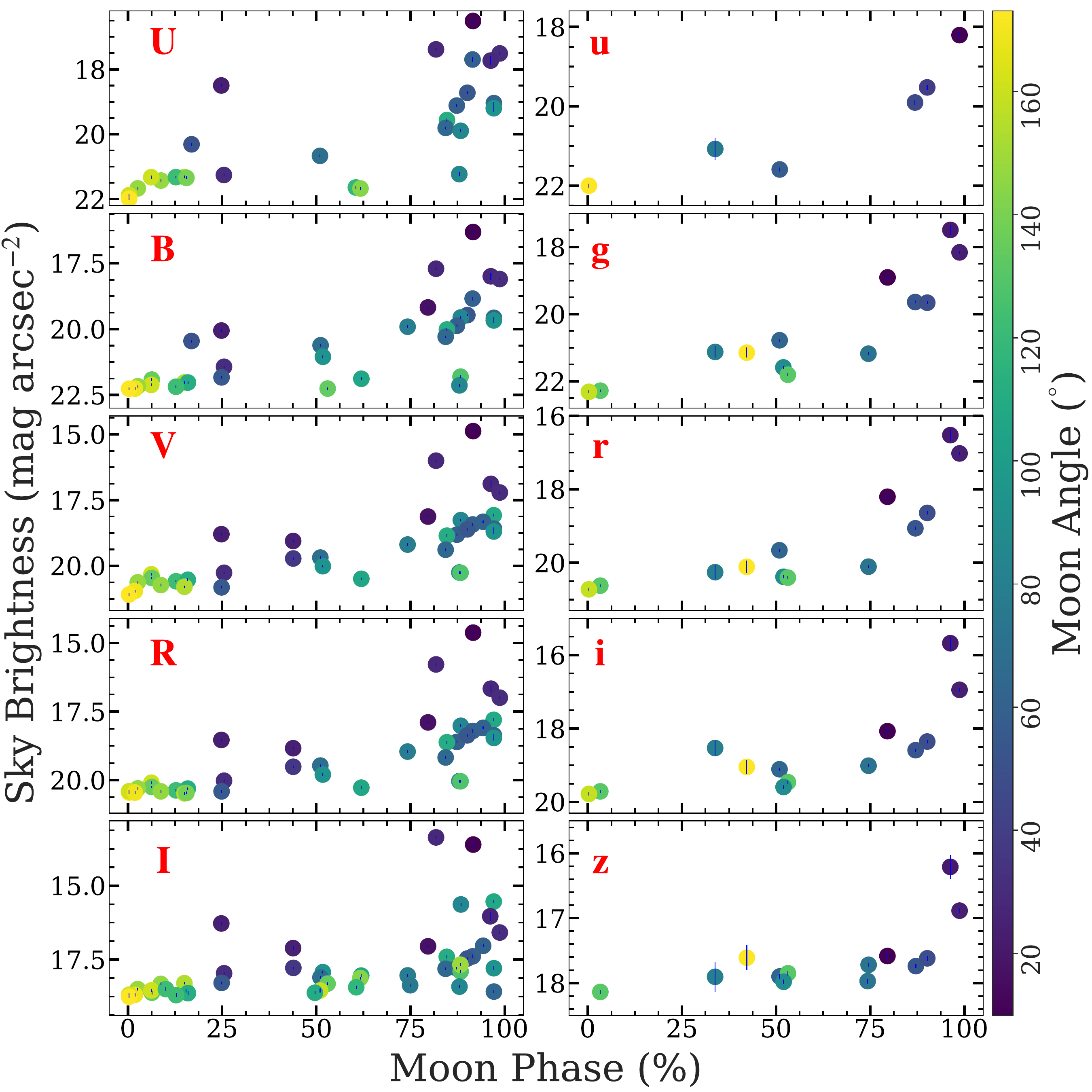}
\caption{Night sky brightness in mag arcsec$^{-2}$ with moon phase (in per cent) and moon angle (in degrees) for Bessell and SDSS filters estimated using the Landolt fields observed from 2016 to 2021 using the 4K$\times$4K CCD Imager on the 3.6m DOT. The values of the night sky brightness are also scaled to the zenith.}
\label{fig:sky_brightness}
\end{figure*}

\begin{table*}[t]
\tabularfont
\caption{Comparison of the night sky brightness in magnitude arcsec$^{-2}$ for Bessell $UBVRI$ filters estimated using the 4K$\times$4K CCD Imager on the 3.6m DOT tabulated along with those published for other good astronomical sites internationally, showing Devasthal site as comparable to other good observing sites.}
\addtolength{\tabcolsep}{10pt}
\begin{tabular}{lccccccc}
\hline
Astronomical sites & $U$ & $B$ & $V$ & $R$ & $I$ & Source\\
\hline 
\hline				
Calar Alto & 22.20 & 22.60 & 21.50 & 20.60 & 18.70 & \cite{Leinert1995} \\
La Silla & --- & 22.81 & 21.75 & 20.08 & 19.5 & \cite{Mattila1996} \\
CTIO & 22.12 & 22.82 & 21.79 & 21.19 & 19.85 & \cite{Krisciunas2007} \\
Paranal & 22.35 & 22.67 & 21.71 & 20.93 & 19.65 & \cite{Patat2008} \\
Hanle & 22.14 & 22.42 & 21.28 & 20.54 & 18.86 & \cite{Stalin2008} \\
\textbf{Devasthal} & \textbf{22.00} & \textbf{22.26} & \textbf{21.09} & \textbf{20.48} & \textbf{18.76} & \textbf{This work} \\
\hline
\label{tab:sky_brightness_comparison}
\end{tabular}
\end{table*}

\section{Sky brightness\label{sec:sky_brightness}}

Even a moonless night sky with an absence of artificial light is also not wholly dark. Its brightness depends on (1) zodiacal light, (2) faint unresolved stars and diffuse galactic light due to atomic processes within our galaxy, (3) diffuse extra-galactic light, and (4) air-glow and aurora \citep{Patat2003, Gill2020}. Out of the above-discussed four sources, the first three sources are independent of the observational site, whereas the fourth one depends on the site and time of observation \citep{Krisciunas1997, Benn1998, Patat2003}. A good astronomical site is characterized by having a low night sky brightness. Sky brightness can be estimated using the formula given by \citealt{Krisciunas1997} (see also \citealt{Stalin2008}):

\begin{equation}\label{eq:sky_brightness}
S = -2.5\ log\frac{C_{sky}}{C_*} + 2.5\ log\frac{E_{sky}}{E_*} + \kappa_\lambda X_* + M_*,
\end{equation}

where, $S$ corresponds to the sky brightness in magnitudes, $C_{sky}$ is the mean sky counts multiplied by the area of the aperture, $C_*$ represents the total count above the sky within the aperture of the standard star, and $E_{sky}$ and $E_*$ are the exposure times corresponding to $C_{sky}$ and $C_{*}$, respectively. $\kappa_\lambda$ is the wavelength-dependent extinction coefficient, $X_*$ is airmass, and $M_*$ is the standard magnitude of the star (adopted from \citealt{Landolt1992}). Using the above equation and the photometric data collected from 2016 through 2021 with the 4K$\times$4K CCD Imager, we estimated the sky brightness values in the Bessell ($UBVRI$) and SDSS ($ugriz$) filters. The sky brightness values given by Equation~\ref{eq:sky_brightness} in magnitudes are converted to mag arcsec$^{-2}$ using the relation:

\begin{equation}\label{eq:sky_brightness_per_arcsec}
I = S + 2.5\ log(A).
\end{equation}

Here, $I$ is the sky brightness in mag arcsec$^{-2}$ and A is the area of the aperture in arcsec$^{2}$ estimated from the given plate scale of the CCD. The sky brightness values are also corrected for zenith distance using the formula:

\begin{equation}\label{eq:sky_brightness_zenith}
\Delta I = -2.5\ log_{10}[(1-f) + fX] + \kappa(X - 1).
\end{equation}

Here, $f$ is a fraction of the total sky brightness generated by airglow, which equals to 0.6 (\citealt{Patat2003}; see also \citealt{Stalin2008}). $X$ is the optical path length along a line of sight which is given as $X$ = (1 - 0.96 sin$^2$ Z)$^{-1/2}$ \citep{Patat2003, Stalin2008}. Hence, the total sky brightness is estimated by simply adding Equations~\ref{eq:sky_brightness_per_arcsec} and \ref{eq:sky_brightness_zenith}.

Data used to estimate the sky brightness values were observed on 32 nights (see Table~\ref{tab:log_table}) distributed over five years (2016--2021) in diverse sky conditions. The sky brightness was estimated for all nights where the standard fields were observed (irrespective of the moon phase and moon angle). However, we chose the data within airmass $<$ 1.5 and time differential from the closest twilight $\Delta$t $>$ 1 hr in the present analysis. Figure~\ref{fig:sky_brightness} shows the sky brightness values with different moon phases and respective moon angles from the observed fields for Bessell ($UBVRI$) and SDSS ($ugriz$) filters. We chose the frame close to the zenith from multiple frames of the same Landolt standard field observed on the same night. It can be seen clearly from Figure~\ref{fig:sky_brightness} that sky brightness is increasing (getting brighter) with moon phase as well and inversely with moon angle. Zenith corrected sky brightness values in the units of mag arcsec$^{-2}$ for the frames with the lowest moon phase and the highest moon angle as estimated from the present analysis are $U$ $\sim$22.00 $\pm$ 0.01, $B$ $\sim$22.26 $\pm$ 0.01, $V$ $\sim$21.09 $\pm$ 0.01, $R$ $\sim$20.48 $\pm$ 0.01, $I$ $\sim$18.76 $\pm$ 0.01, $u$ $\sim$22.00 $\pm$ 0.02, $g$ $\sim$22.31 $\pm$ 0.01, $r$ $\sim$20.72 $\pm$ 0.01, $i$ $\sim$19.78 $\pm$ 0.01, $z$ $\sim$18.14 $\pm$ 0.01.

The night sky brightness estimated for the Devasthal site in $UBVRI$ filters presented in this study are also compared with corresponding values from some of the other well known astronomical sites: Calar Alto, La Silla, CTIO, Paranal, and Hanle (see Table~\ref{tab:sky_brightness_comparison}). The night sky brightness for the Devasthal site appears close to those of other good astronomical sites discussed, making Devasthal one of the darkest astronomical observing sites.

\section{Results and Conclusion\label{sec:result}}

In this work, we present the characterization results of the 4K$\times$4K CCD Imager (mounted at the axial port of the 3.6m DOT) by verifying bias stability and gain/RN parameters in all the combinations of the readout speeds (100 kHz, 500 kHz, and 1 MHz) based on the data acquired recently in early 2021. Using the data observed spanning nearly five years, we have also constrained the values of photometric transformation coefficients and photometric precision for the 4K$\times$4K CCD Imager. Using the suitable data sets, measurements of atmospheric extinction coefficients and night sky brightness values are also performed for the Devasthal site. The present characterization of the 4K$\times$4K CCD Imager and the photometric calibration results will be helpful in calibrating observations of various astronomical objects, both Galactic and extra-galactic. The key outcomes of this study are summarized below:

$\bullet$ The distribution of FWHM values obtained for $V$-band data during 2016-2021 shows a median of 1.24 arcsec and sub-arcsec values for $\sim$10\% of the observed sample. The best FWHM value obtained was about 0.43 arcsec (in SDSS $r$-band) during the cycle 2020A (on 2020 March 17).

$\bullet$ Based on the data taken in 2020, we measured the new transmission curves of all the ten broadband filters and compared them with the quantum efficiency curve of the STA CCD. Recent observations also establish the sustained bias stability of the Imager as per the estimated LN2 hold time ($\sim$14 hours) or for the time duration with CCD temperatures being --120$^\circ C$ or comparable.

$\bullet$ The gain and $RN$ values are verified for all the possible combinations of readout speeds (100 kHz, 500 kHz, and 1 MHz) and theoretical gain values (1, 2, 3, 5, and 10 e$^-$/ADU). The measured values of gain and $RN$ are consistent with the theoretical ones with discrepancies of a few \% for a single readout mode.

$\bullet$ The estimated values of extinction coefficients in mag airmass$^{-1}$ based on the data obtained in 2021 are: $U$ = 0.43 $\pm$ 0.01, $B$ = 0.21 $\pm$ 0.01, $V$ = 0.12 $\pm$ 0.01, $R$ = 0.07 $\pm$ 0.01, $I$ = 0.03 $\pm$ 0.01, $u$ = 0.48 $\pm$ 0.03, $g$ = 0.20 $\pm$ 0.01, $r$ = 0.09 $\pm$ 0.01, $i$ = 0.08 $\pm$ 0.02, and $z$ = 0.05 $\pm$ 0.02, and these are consistent with those were calculated theoretically. The extinction coefficients for the Devasthal site are in the range of those reported for other well-known astronomical sites discussed in this work. A signature of seasonal variation of the extinction values is also evident from our analysis as reported in previous works published for nearby sites.

$\bullet$ The mean values of color coefficients for photometric calibration are found to be $\alpha_0$ = 1.09 $\pm$ 0.04, $\alpha_1$ = 1.01 $\pm$ 0.03, $\alpha_2$ = 1.01 $\pm$ 0.03, $\alpha_3$ = 0.84 $\pm$ 0.01, $\alpha_4$ = 0.91 $\pm$ 0.02, $\alpha_5$ = $-$0.04 $\pm$ 0.02, $\alpha_6$ = $-$0.04 $\pm$ 0.03, $\alpha_7$ = 0.02 $\pm$ 0.02, $\alpha_8$ = 0.05 $\pm$ 0.03, and $\alpha_9$ = $-$0.02 $\pm$ 0.03. Based on our analysis, we do not find any noticeable temporal evolution and the values of photometric transformation coefficients during 2016 to 2021 are within 1$\sigma$ limit.

$\bullet$ difference between the Landolt and the present photometry in $UBVRI$ filters are consistent within 1$\sigma$ for a brightness range of 12-16 mag in the $V$-band. Also, the limiting magnitude values ($V$ =19.5 $\pm$ 0.0.05 mag for a 40 sec exposure / $V$=21.5 $\pm$ 0.05 mag for a 360 sec exposure time) follow the simulated ones as per criteria discussed in \textcolor{blue}{P18}\href{https://ui.adsabs.harvard.edu/abs/2018BSRSL..87...42P/abstract}.

$\bullet$ Zenith corrected night sky brightness values for the Devasthal site in the units of mag arcsec$^{-2}$ for the moon-less nights are $U$ $\sim$22.00 $\pm$ 0.01, $B$ $\sim$22.26 $\pm$ 0.01, $V$ $\sim$21.09 $\pm$ 0.01, $R$ $\sim$20.48 $\pm$ 0.01, $I$ $\sim$18.76 $\pm$ 0.01, $u$ $\sim$22.00 $\pm$ 0.02, $g$ $\sim$22.31 $\pm$ 0.01, $r$ $\sim$20.72 $\pm$ 0.01, $i$ $\sim$19.78 $\pm$ 0.01, and $z$ $\sim$18.14 $\pm$ 0.01. The values of sky brightness are comparable to the other well-known sites for optical/NIR observations.\\

\appendix
\setcounter{table}{0}
\renewcommand{\thetable}{A\arabic{table}}
\setcounter{figure}{0}
\renewcommand{\thefigure}{A\arabic{figure}}

\section{} 
\begin{figure}
\centering
\includegraphics[angle=0,scale=0.4]{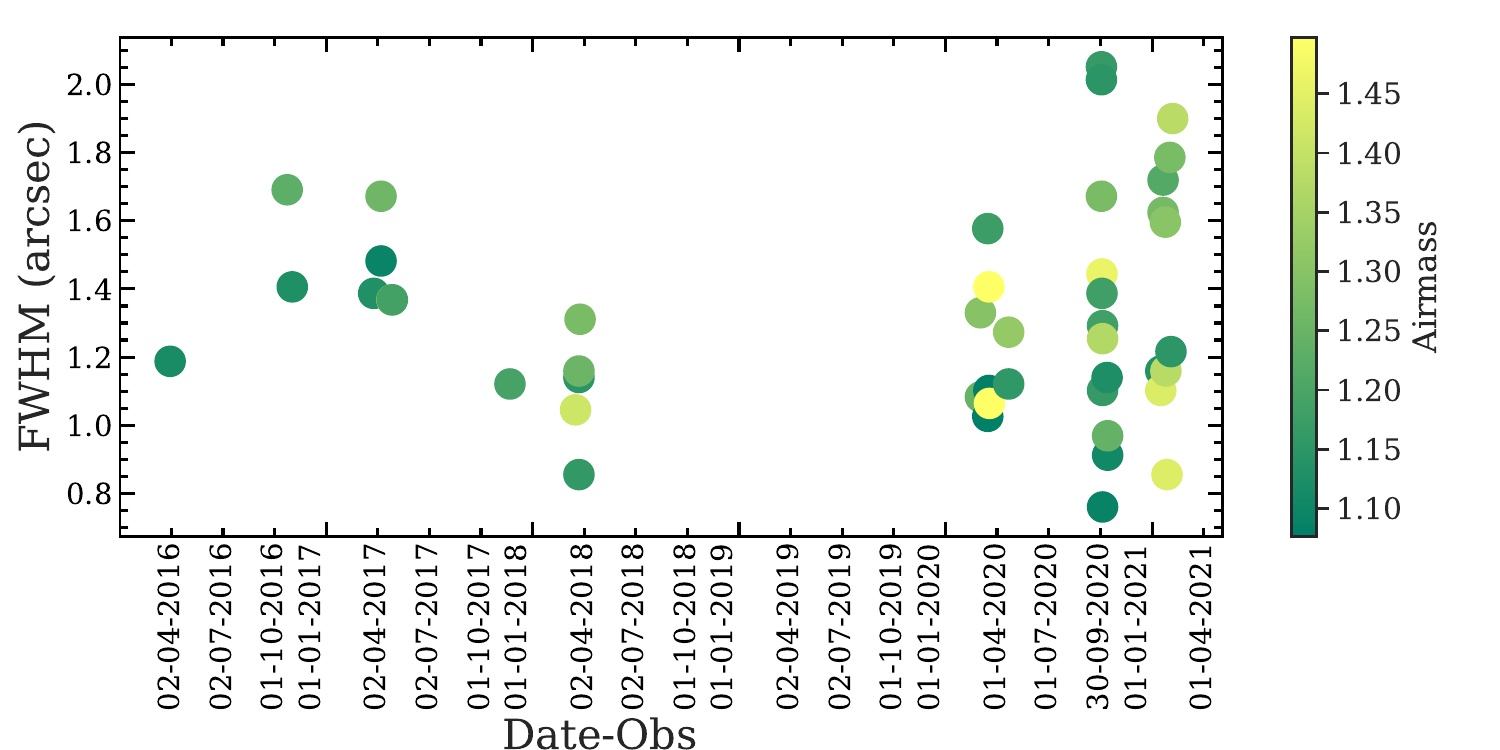}
\caption{FWHM values (in arcsec) estimated using $V$-band imaging data of the standard fields \citep{Landolt1992} discussed in Table~\ref{tab:log_table} using the 4K$\times$4K CCD Imager mounted at the axial port of the 3.6m DOT during 2016--2021. For nearly 10\% of nights, the FWHM values are $<$ 1 arcsec for $V$-band.}
\label{fig:fwhm_three_cycle}
\end{figure}

\begin{figure*}[!b]
\centering
\includegraphics[angle=0,scale=0.29]{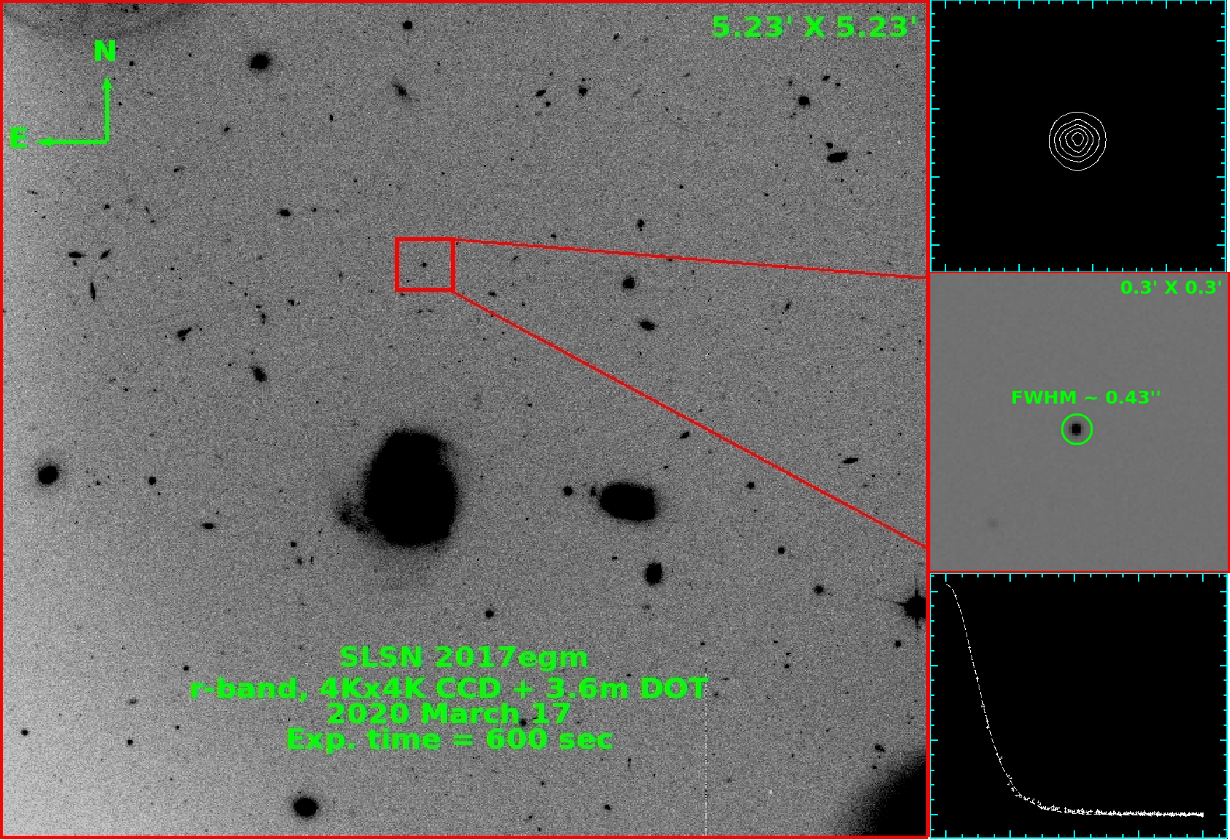}
\caption{The field of SN~2017egm observed in $r$-band on 2020 March 17, for an exposure time of 600s is presented. This image clearly detects many galaxies in the field as faint as $\sim$22 mag and several faint point sources as expected with this CCD Imager/3.6m DOT. The FWHM of the stellar profile (in red box) was observed as $\sim$0.43 arcsec. The contour and radial profile of the stellar source are also shown.}
\label{fig:fwhm_best}
\end{figure*}

\begin{figure*}
\centering
\includegraphics[angle=0,scale=0.375]{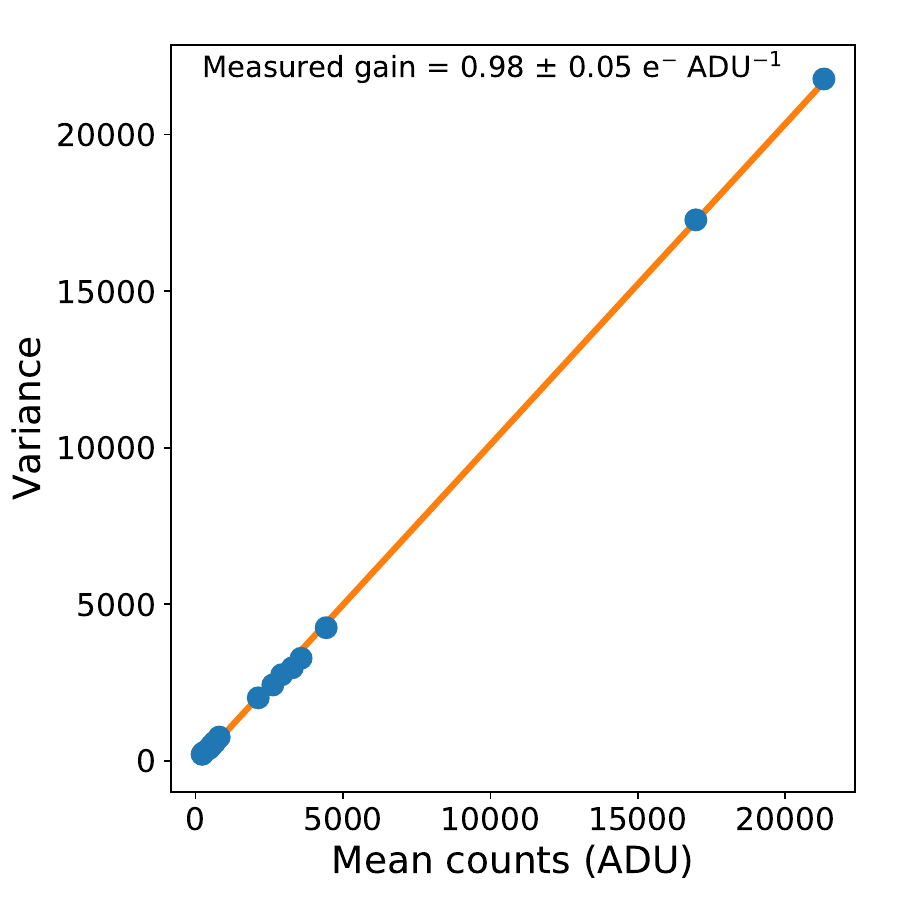}
\includegraphics[angle=0,scale=0.375]{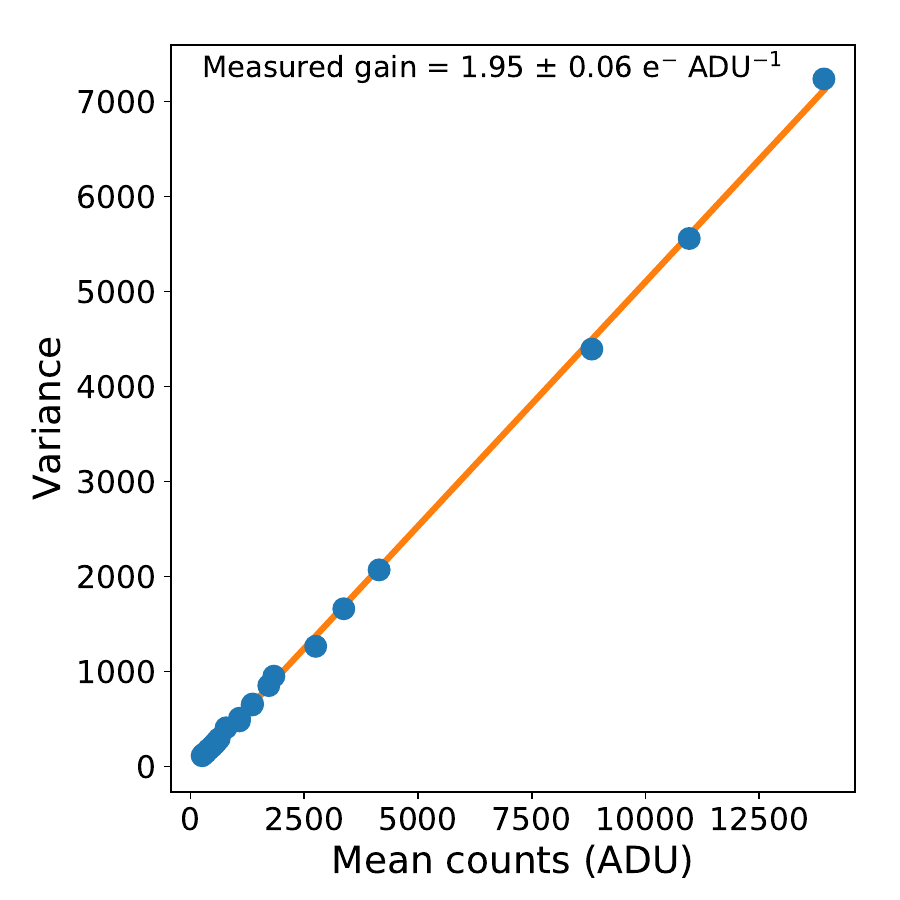}
\includegraphics[angle=0,scale=0.375]{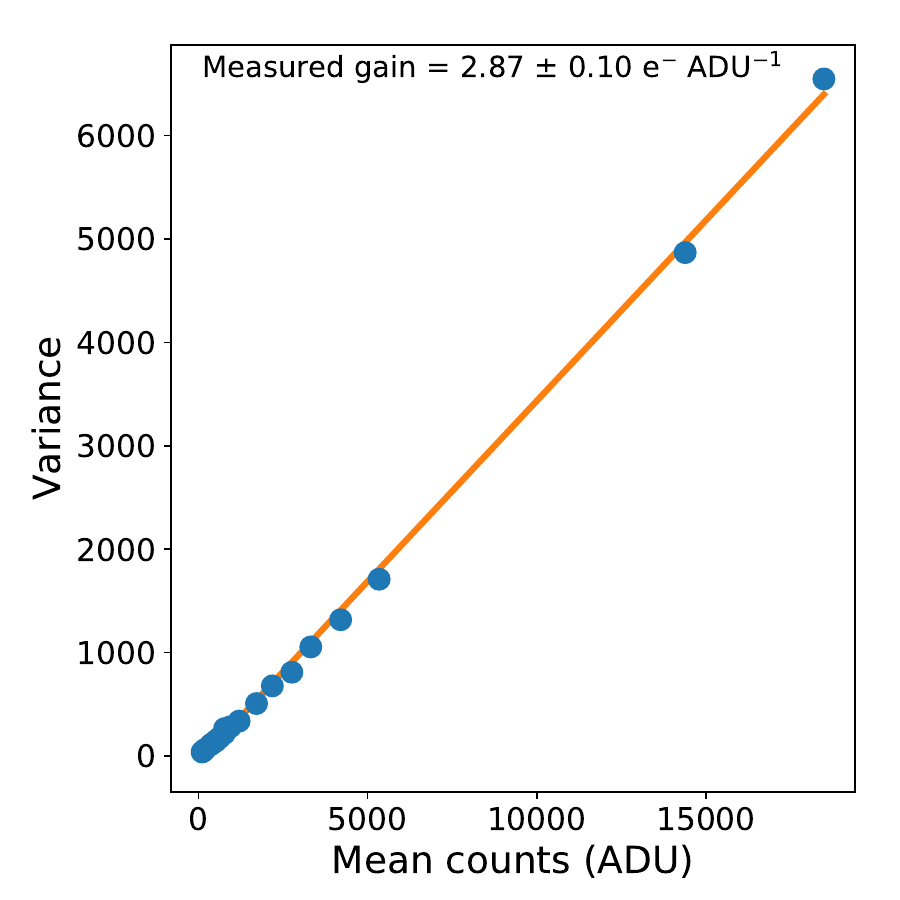}
\includegraphics[angle=0,scale=0.375]{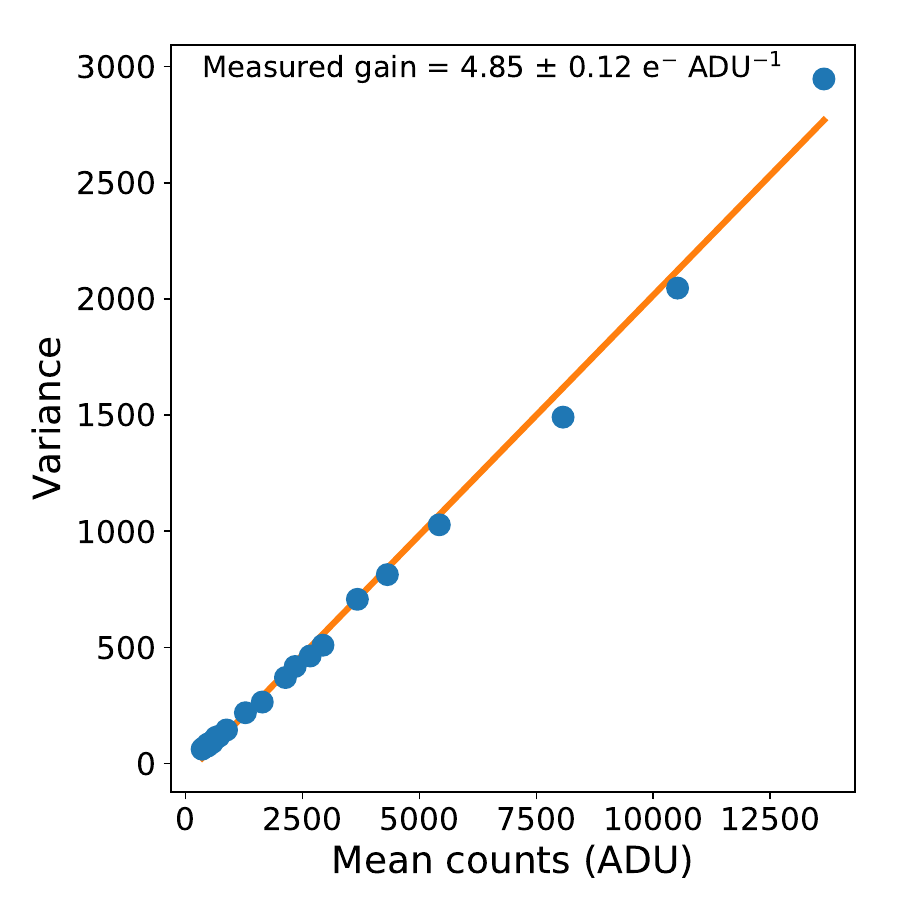}
\includegraphics[angle=0,scale=0.375]{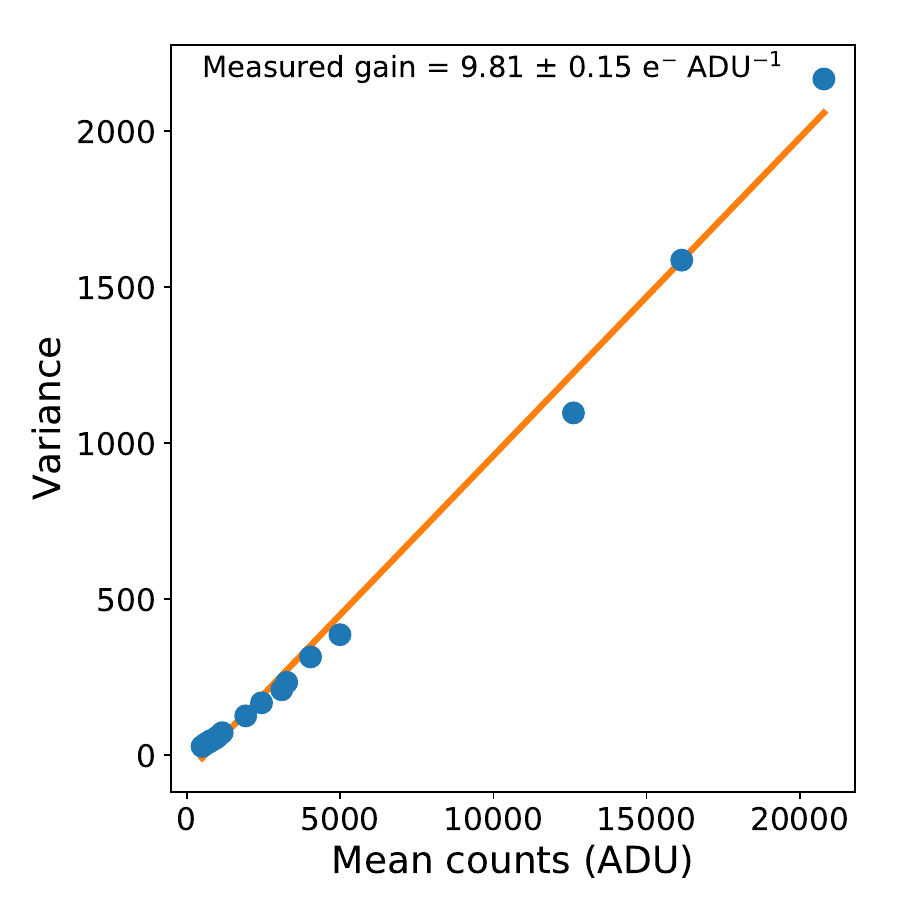}
\caption{PTCs of the 4K$\times$4K CCD Imager taken at readout speed of 100 kHz and for all available gain values. The measured gain values are $\sim$0.98$\pm$0.05, 1.95$\pm$0.06, 2.87$\pm$0.10, 4.85$\pm$0.12, and 9.81$\pm$0.15 e$^-$/ADU for theoretical gain values of 1, 2, 3, 5, and 10 e$^-$/ADU, respectively.}
\label{fig:gain_ptc}
\end{figure*}

The observations log of all the Landolt standard fields used in this study is tabulated in Table~\ref{tab:log_table}. The FWHM (in arcsec; for $V$-band) values for all these Landolt fields (having an airmass range of $\sim$1 to 1.5) were also estimated (see Figure~\ref{fig:fwhm_three_cycle}) using IRAF/$imexa$. Figure~\ref{fig:fwhm_three_cycle} shows the distribution of measured FWHM values in the $V$-band for present observations distributed over 32 nights during a span of 5 years. For nearly 10\% of the observed sample, $<$1 arcsec FWHM values were observed. The best FWHM obtained for a point source was about 0.43 arcsec on 2020 March 17 (see a $r$-band stellar image in Figure~\ref{fig:fwhm_best}) along with many faint galaxies detected in the field. We have also shown the contour and radial profile of the stellar image in Figure~\ref{fig:fwhm_best}, respectively. The FWHM of the stellar profile is 2.26 pixels which converts to $\sim$0.43 arcsec using the plate scale of the CCD (0.19 arcsec per pixel for 2$\times$2 binning) on the sky. As an example, Figure~\ref{fig:gain_ptc} represents the PTCs for obtaining the gain values for readout speed of 100 kHz and theoretical gain values of 1, 2, 3, 5, and 10 e$^-$/ADU. In a similar way, the gain values are verified for readout speeds of 500 kHz and 1 MHz; those are tabulated in Table~\ref{tab:gain_rn}.

\onecolumn
\begin{center}
\begin{small}
\begin{longtable}{|p{2.0cm}|p{2.0cm}|p{2.0cm}|p{2.0cm}|p{2.0cm}|p{2.0cm}|p{2.0cm}|}
\caption{Log of Landolt standard fields \citep{Landolt1992} used in this study observed in Bessell $UBVRI$ and SDSS $ugriz$ filters from 2016 to 2021.} \label{tab:log_table} \\
\hline \multicolumn{1}{|c|}{Date} & \multicolumn{1}{c|}{Standard} & \multicolumn{1}{c|}{\textit{U}} & \multicolumn{1}{c|}{\textit{B}}& \multicolumn{1}{c|}{\textit{V}}& \multicolumn{1}{c|}{\textit{R}}& \multicolumn{1}{c|}{\textit{I}}\\ \hline 
\endfirsthead

\multicolumn{7}{c}%
{{\bfseries \tablename\ \thetable{} -- continued from previous page}} \\
\hline \multicolumn{1}{|c|}{Date} & \multicolumn{1}{c|}{Standard} & \multicolumn{1}{c|}{\textit{U}} & \multicolumn{1}{c|}{\textit{B}}& \multicolumn{1}{c|}{\textit{V}}& \multicolumn{1}{c|}{\textit{R}}& \multicolumn{1}{c|}{\textit{I}}\\ \hline 
\endhead

\hline \multicolumn{7}{|c|}{{Continued on next page}} \\ \hline
\endfoot

\hline
\endlastfoot

2016-03-31 &    PG0918    & 200s$\times$4   &    200s$\times$3    &    200s$\times$2    &    200s$\times$2    &    200s$\times$2\\
2016-10-24 &    PG2213    & 200s$\times$2   &    200s$\times$2    &    200s$\times$2    &    200s$\times$2    &    200s$\times$6 \\
2016-11-02 &    PG0231    & 200s$\times$3   &    100s$\times$3    &    100s$\times$3    &    100s$\times$3    &    100s$\times$3\\
2017-03-26 &    PG~1633    & 60xs12   &    30s$\times$12    &    20s$\times$12    &    20s$\times$12    &    20s$\times$12\\
2017-04-08 &    PG~1525    & ---      &    ---       &    10s$\times$3     &    10s$\times$3     &    10s$\times$3 \\
2017-04-08 &    PG~1528    & ---      &    ---       &    6s$\times$4      &    5s$\times$6      &    5s$\times$4 \\
2017-04-16 &    PG0918    & 150s$\times$9   &    100s$\times$4    &    100s$\times$5    &    40s$\times$3     &    40s$\times$3 \\
           &              &          &    150s$\times$2    &    60s$\times$1     &    60s$\times$2     &    60s$\times$2 \\
2017-04-16 &    SA~104     & 150s$\times$11  &    150s$\times$5    &    60s$\times$5     &    60s$\times$6     &    60s$\times$7  \\
2017-04-16 &    PG~1633    & 200s$\times$3   &    100s$\times$2    &    50s$\times$2     &    20s$\times$3     &    20s$\times$4  \\
2017-04-27 &    PG~1323    & 100s$\times$1   &    60s$\times$1     &    30s$\times$1     &    20s$\times$1     &    20s$\times$2  \\
2017-04-28 &    SA~110     & 100s$\times$1   &    50s$\times$2     &    25s$\times$2     &    25s$\times$2     &    30s$\times$2  \\
2017-11-21 &    PG0231    & 100s$\times$3   &    50s$\times$2     &    30s$\times$2     &    30s$\times$2     &    20s$\times$3  \\
2017-11-22 &    PG0231    & 200s$\times$3   &    60s$\times$3     &    30s$\times$3     &    20s$\times$3     &    20s$\times$3  \\
2018-03-18 &    PG~1323    & ---      &    100s$\times$2    &    40s$\times$2     &    30s$\times$2     &    30s$\times$2  \\
2018-03-24 &    PG~1047    & 60s$\times$2    &    15s$\times$2     &    20s$\times$2     &    10s$\times$2     &    10s$\times$2  \\
2018-03-24 &    PG~1525    & 80s$\times$1    &    40s$\times$1     &    20s$\times$1     &    10s$\times$1     &    10s$\times$1  \\
2018-03-24 &    SA~104     & 150s$\times$3   &    30s$\times$2     &    20s$\times$2     &    20s$\times$2     &    15s$\times$2  \\
2018-03-26 &    PG~1323    & ---      &    80s$\times$13    &    60s$\times$14    &    20s$\times$12    &    20s$\times$12  \\
2020-03-03 &    PG~1323    & 100s$\times$3   &    60s$\times$3     &    30s$\times$3     &    30s$\times$3     &    30s$\times$3  \\
2020-03-03 &    PG~1525    & 60s$\times$2    &    30s$\times$2     &    30s$\times$2     &    30s$\times$2     &    30s$\times$2   \\
2020-03-16 &    PG~1657    & ---      &    ---       &    30s$\times$2     &    30s$\times$2     &    30s$\times$2   \\
2020-03-16 &    STD~109    & ---      &    30s$\times$2     &    30s$\times$2     &    30s$\times$2     &    30s$\times$2   \\
2020-03-18 &    PG~1657    & 60s$\times$2    &    40s$\times$2     &    30s$\times$2     &    30s$\times$2     &    30s$\times$2  \\
2020-03-18 &    STD~111    & 60s$\times$2    &    40s$\times$2     &    30s$\times$2     &    30s$\times$2     &    30s$\times$2  \\
2020-03-19 &    STD~109    & 40s$\times$2    &    30s$\times$2     &    ---     &    ---     &    ---  \\
2020-03-19 &    STD~110    & 40s$\times$2    &    30s$\times$2     &    20s$\times$2     &    20s$\times$2     &    20s$\times$2  \\
2020-04-22 &    PG~1323    & 40s$\times$2    &    30s$\times$2     &    30s$\times$2     &    20s$\times$2     &    15s$\times$2   \\
2020-04-22 &    STD~104    & 30s$\times$2    &    30s$\times$2     &    20s$\times$2     &    20s$\times$2     &    20s$\times$2   \\
2020-10-03 &    STD~95     & 120s$\times$15  &    120s$\times$10   &    120s$\times$7    &    120s$\times$7    &    120s$\times$7  \\
2020-10-03 &    STD~110    & 60s$\times$5    &    50s$\times$3     &    40s$\times$2     &    30s$\times$2     &    30s$\times$2   \\
2020-10-03 &    STD~111    & 60s$\times$5    &    40s$\times$2     &    30s$\times$2     &    20s$\times$2     &    20s$\times$2   \\
2020-10-03 &    STD~113    & 50s$\times$2    &    40s$\times$2     &    30s$\times$2     &    20s$\times$2     &    20s$\times$2   \\
2020-10-04 &    PG0231    & 150s$\times$2   &    120s$\times$2    &    90s$\times$2     &    60s$\times$2     &    60s$\times$2   \\
2020-10-04 &    STD~98     & 180s$\times$1   &    150s$\times$1    &    120s$\times$1    &    120s$\times$1    &    120s$\times$1  \\
2020-10-05 &    PG~1633    & 90s$\times$4    &    60s$\times$4     &    60s$\times$4     &    30s$\times$4     &    30s$\times$4   \\
2020-10-05 &    PG2332    & 120s$\times$7   &    120s$\times$4    &    120s$\times$6    &    120s$\times$3    &    120s$\times$6  \\
2020-10-05 &    STD~110    & 90s$\times$2    &    60s$\times$2     &    60s$\times$2     &    30s$\times$2     &    30s$\times$2   \\
2020-10-05 &    STD~113  & 90s$\times$4    &    90s$\times$4     &    90s$\times$4     &    60s$\times$3     &    60s$\times$4   \\
2020-10-13 &    PG0231    & 120s$\times$2   &    60s$\times$2     &    30s$\times$2     &    30s$\times$2     &    30s$\times$4   \\
2020-10-14 &    PG0231    & 60s$\times$2    &    30s$\times$2     &    20s$\times$2     &    30s$\times$2     &    30s$\times$3   \\
2020-10-14 &    STD~92     & 150s$\times$2   &    120s$\times$2    &    90s$\times$2     &    40s$\times$5     &    50s$\times$2   \\
2021-01-16 &    PG~1323    & 40s$\times$2    &    30s$\times$2     &    20s$\times$2     &    15s$\times$2     &    15s$\times$2   \\
2021-01-16 &    PG0918    & 40s$\times$2    &    20s$\times$2     &    15s$\times$2     &    10s$\times$2     &    10s$\times$3   \\
2021-01-20 &    PG0918    & ---      &    120s$\times$4    &    100s$\times$4    &    100s$\times$4    &    70s$\times$4   \\
2021-01-20 &    STD~104    & ---      &    250s$\times$3    &    150s$\times$3    &    100s$\times$3    &    90s$\times$3   \\
2021-01-24 &    PG~1323    & 20s$\times$3    &    10s$\times$3     &    5s$\times$3      &    5s$\times$3      &    5s$\times$9    \\
2021-01-24 &    PG0918    & 20s$\times$3    &    10s$\times$3     &    8s$\times$3      &    5s$\times$3      &    5s$\times$11   \\
2021-01-24 &    STD~104    & 120s$\times$2   &    60s$\times$2     &    40s$\times$2     &    30s$\times$2     &    20s$\times$6   \\
2021-01-25 &    PG0918    & 120s$\times$11  &    60s$\times$11    &    30s$\times$11    &    30s$\times$11    &    30s$\times$11  \\
2021-01-27 &    PG0921    & 120s$\times$14  &    60s$\times$14    &    30s$\times$15    &    30s$\times$15    &    30s$\times$15  \\
2021-02-01 &    PG~1323    & ---      &    60s$\times$1     &    30s$\times$1     &    30s$\times$2     &    30s$\times$1 \\
2021-02-03 &    STD~98     & ---      &    60s$\times$3     &    60s$\times$5     &    60s$\times$4     &    40s$\times$5  \\
2021-02-06 &    PG~1657    & 150s$\times$19  &    60s$\times$19    &    20s$\times$20    &    20s$\times$20    &    20s$\times$20  \\
2021-02-07 &    PG0918    & 30s$\times$3    &    15s$\times$2     &    10s$\times$3     &    10s$\times$3     &    10s$\times$3   \\
2021-02-07 &    PG~1633    & 120s$\times$1   &    60s$\times$1     &    30s$\times$1     &    20s$\times$1     &    20s$\times$1   \\
2021-02-07 &    PG~1657    & 120s$\times$6   &    60s$\times$7     &    20s$\times$7     &    20s$\times$7     &    20s$\times$7   \\
\hline
  Date         &    Field          & $u$      &    $g$       &      $r$     &      $i$     &     $z$  \\
\hline
2017-03-26 &    PG~1633    & 60s$\times$2   &    20s$\times$2    &    20s$\times$2    &    20s$\times$2    &    20s$\times$2 \\
2017-04-16 &    SA~104     & 150s$\times$7   &   150s$\times$3    &   60s$\times$3    &    60s$\times$3    &    60s$\times$3 \\
           &              & 100s$\times$3   &   80s$\times$3    &    50s$\times$3    &    50s$\times$3    &    50s$\times$2 \\
2017-04-27 &    STD~110     & 100s$\times$1   &   20s$\times$2    &   10s$\times$2    &    5s$\times$2    &    5s$\times$2 \\           
2018-03-24 &    PG~1047    & 60s$\times$2    &    30s$\times$2     &    10s$\times$3     &    10s$\times$2     &    10s$\times$2 \\
2018-03-24 &    PG~1525    & 100s$\times$1   &    60s$\times$1     &    20s$\times$1     &    15s$\times$1     &    15s$\times$1  \\
2018-03-24 &    STD~104     & 150s$\times$2   &    80s$\times$3     &    20s$\times$2     &    20s$\times$2     &    20s$\times$2   \\
2018-03-26 &    PG~1323    & 120s$\times$2  &    ---       &    ---       &    ---       &    20s$\times$2  \\
2020-03-18 &    PLUS2     & 5s$\times$2  &    5s$\times$2    &    3s$\times$2    &    3s$\times$2   &    3s$\times$2 \\
2020-04-22 &    STD~104    & 30s$\times$2  &    30s$\times$2    &    20s$\times$2    &    20s$\times$2   &    --- \\
2020-10-03 &    STD~95     & 120s$\times$2  &    120s$\times$2    &    120s$\times$2    &    120s$\times$2   &    120s$\times$2 \\
2020-10-05 &    PG~1633    & ---      &    ---       &    ---       &    30s$\times$2     &    30s$\times$2 \\
2020-10-05 &    PG2332    & 120s$\times$4   &    120s$\times$4    &    120s$\times$4    &    120s$\times$4    &    120s$\times$4  \\
2021-01-20 &    PG~1323    & 15s$\times$2    &    10s$\times$2     &    30s$\times$2     &    10s$\times$2     &    15s$\times$2s \\
2021-01-25 &    PG0918    & 120s$\times$9   &    60s$\times$10    &    30s$\times$10    &    30s$\times$10    &    30s$\times$10  \\
2021-01-27 &    PG0921    & 120s$\times$2  &    60s$\times$2    &    30s$\times$2    &    30s$\times$2    &    30s$\times$2  \\
2021-02-01 &    PG~1323    & 60s$\times$3    &    60s$\times$3     &    15s$\times$1     &    10s$\times$1     &    10s$\times$1 \\
\end{longtable}
\end{small}
\end{center}

\twocolumn

\section*{Acknowledgements}
This study uses data observed using the 4K$\times$4K CCD Imager mounted at the axial port of the 3.6m DOT from 2016 to 2021. The data of the Landolt standards in this study obtained by the Imager team were used for tests and calibration purposes. The authors of this paper are highly thankful to the observing staff and the observing assistants for their support during observations with the 3.6m DOT. AK, SBP, and RKSY also acknowledge 3.6m DOT proposals DOT-2020-C2-P42, DOT-2021-C1-P55, and instrument verification nights in obtaining the data used in this study. The authors of this paper thank the anonymous referee for his/her detailed and constructive comments to improve the overall analysis. S.B.P. also acknowledge BRICS grant DST/\\IMRCD/BRICS/Pilotcall/ProFCheap/2017(G) for the present work.

\bibliographystyle{apj}
\bibliography{reference}

\end{document}